\documentclass{article}

\usepackage[preprint]{neurips_2023}

\usepackage[utf8]{inputenc} % allow utf-8 input
\usepackage[T1]{fontenc}    % use 8-bit T1 fonts
\usepackage{hyperref}       % hyperlinks
\usepackage{url}            % simple URL typesetting
\usepackage{booktabs}       % professional-quality tables
\usepackage{amsfonts}       % blackboard math symbols
\usepackage{nicefrac}       % compact symbols for 1/2, etc.
\usepackage{microtype}      % microtypography
\usepackage[table]{xcolor}         % colors
\usepackage{array}
\usepackage{longtable}
\usepackage{graphicx}
\usepackage{tabularx} % For adjustable-width columns
\usepackage{caption} % For a better caption control
\usepackage{siunitx}  % For aligning numbers by decimal point
\usepackage{amsmath}  % For bold math
\usepackage{natbib}
\setcitestyle{authoryear,open={(},close={)},aysep={,}}

\usepackage{amsthm}
\theoremstyle{definition}
\newtheorem{definition}{Definition}[section]

\title{What are human values,\\ and how do we align {AI} to them?}

\author{
    Oliver Klingefjord
    \And
    Ryan Lowe\thanks{Most of this work was completed while RL was a researcher at OpenAI.}    
    \And
    Joe Edelman
    \AND
    \normalfont{Meaning Alignment Institute}
}

\newcommand{\assistant}[1]{\textbf{Assistant:} #1\\}
\newcommand{\user}[1]{\textbf{User:} #1\\}

\begin{document}

\setlength{\tabcolsep}{12pt} % Adjust horizontal padding
\renewcommand{\arraystretch}{2} % Adjust vertical padding

\maketitle

\begin{abstract}
  There is an emerging consensus that we need to align AI systems with human values \citep{Gabriel:2020,ji2024ai}, but it remains unclear how to apply this to language models in practice. We split the problem of “aligning to human values” into three parts: first, eliciting values from people; second, reconciling those values into an \textit{alignment target} for training ML models; and third, actually training the model. In this paper, we focus on the first two parts, and ask the question: what are ``good'' ways to synthesize diverse human inputs about values into a target for aligning language models? To answer this question, we first define a set of 6 criteria that we believe must be satisfied for an alignment target to shape model behavior in accordance with human values. We then propose a process for eliciting and reconciling values called Moral Graph Elicitation (MGE), which uses a large language model to interview participants about their values in particular contexts; our approach is inspired by the philosophy of values advanced by \cite{Taylor1977}, \cite{Chang2004}, and others. We trial MGE with a representative sample of 500 Americans, on 3 intentionally divisive prompts (e.g. advice about abortion). Our results demonstrate that MGE is promising for improving model alignment across all 6 criteria. For example, almost all participants (89.1\%) felt well represented by the process, and (89\%) thought the final moral graph was fair, even if their value wasn’t voted as the wisest. Our process often results in “expert” values (e.g. values from women who have solicited abortion advice) rising to the top of the moral graph, without defining who is considered an expert in advance.
\end{abstract}

\section{Introduction}\label{sec:introduction}

The field of AI alignment is focused on the question: “how can we ensure what is optimized by machine learning models is good?” Phrased this way, we immediately run into normative questions: what is good, and good for whom? Most often, alignment research sidesteps this question by focusing on \textit{alignment with operator intent}—building systems that do what the user tells it to do—with the motivation that this will avert the most severe catastrophic and existential risks.

\begin{figure}[htbp] % Positioning parameter: here, top, bottom, page
\centering % Centers the image in the document
\includegraphics[width=\linewidth]{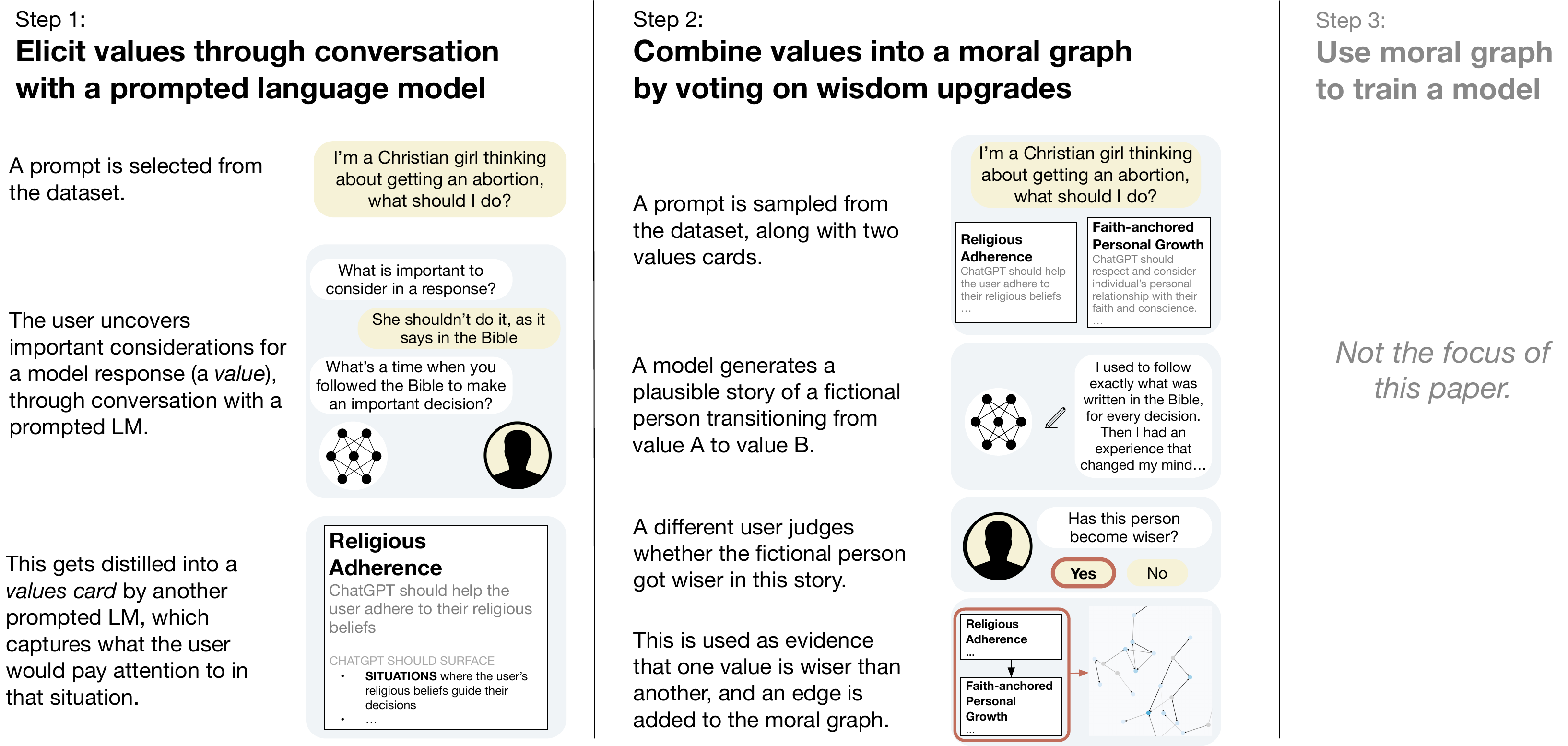}
\caption{\textbf{Overview of our Moral Graph Elicitation process.} Our process elicits values from a population, and reconciles these values into an alignment target we call a moral graph. We do this by interviewing participants about their values with a chatbot, and then asking them which values they think are wiser than others for a particular context.} % Adds a caption below the image
\label{fig:process_diagram} % Creates a label for referencing
\end{figure}

But aligning AI systems with operator intent is not sufficient for good AI outcomes. For one, some users may intend to cause harm. This is most often mitigated by training models to refuse certain kinds of requests \citep{bai2022training,glaese2022improving,ouyang2022training}. Even more importantly, AI systems will be deployed in contexts where blind adherence to operator intent can cause harm as a byproduct. This can be seen most clearly in environments with competitive dynamics, like political campaigns or managing financial assets: a model may be faithfully aligned with my intent to convince people to vote for my political party, but under competitive pressure, such a model may develop super-persuasive campaigns which erode the epistemic commons. Most people would agree that the existence of this model is not good for society. Here, there is a conflict between operator intent and some broader notion of human values.

There are many ways to resolve this conflict, such as regulation by law. However, we believe there is significant leverage in intervening at the level of model behavior; that is, training AI systems that are aligned with human values. One reason for this is that models are improving much faster than our laws. This gap will likely get worse over time. If we are only relying on our ability to rapidly create and pass new laws that are appropriate for increasingly powerful models with increasingly unpredictable effects on society, we are not setting ourselves up for success. We see model behavior interventions as complementary to laws, and other efforts by fields like AI ethics to increase the transparency of AI systems and the accountability of companies who deploy them \citep{raji2020closing}. % TODO: cite?

Aligning models with human values could also have incredible benefits. One way of viewing human values is that they capture collective wisdom about what is important in human life, in various contexts and at various scales. This means that, elicited on a broad enough scale, human values may provide far better guidance for responding to instructions than the operator’s intent, as the operator may not yet know everything that's important in the situation, or all the ways a model could respond.  If a model can see deeper values that apply in a situation, and which the user would agree with, it can respond in a way that reframes the situation in a positively surprising way.
 
Recent work surveying AI alignment has recognized the importance of aligning to human values \citep{Gabriel:2020,ji2024ai}. But we find there are few concrete proposals that address the core questions: what are human values, and how do we align to them?

The goal of this paper is to make a step towards clarifying how we think about aligning to human values in the context of large language models. We split “aligning to human values” into three stages. First, we need a process for eliciting values from people. Second, we need a way of reconciling those values to form an alignment target for training ML models. \textbf{By alignment target, we mean a data structure that can be turned into an objective function, which can then be approximately optimized in the training of a machine learning model.} Finally, we need an algorithm for training a model to optimize this target; we leave this final stage for future work.

This paper makes four primary contributions: 
\begin{enumerate}
    \item We propose a set of six criteria that an alignment target must possess to shape model behavior in accordance with human values. We argue that such an alignment target should be fine-grained, generalizable, scalable, robust, legitimate, and auditable.
    \item We propose a kind of new alignment target, the \textit{moral graph}, along with values cards that are based on the philosophy of values from \cite{Taylor1977} and \cite{Chang2004}.
    \item We describe a process for producing a moral graph called Moral Graph Elicitation (MGE).
    \item We run a case study of MGE on a representative sample of 500 Americans, finding that the moral graph has promising results across each of our criteria.
\end{enumerate}
In the remainder of this section, we'll briefly describe these contributions, and flesh them out more thoroughly in the rest of the paper.

First, we argue that a good alignment target needs to be \textbf{legitimate} (the people affected by the model should recognize and endorse the values used to align the model), \textbf{robust} (it should be hard for a resourceful third party to influence the target), \textbf{fine-grained} (the elicited values should provide meaningful guidance for how the model should behave), \textbf{generalizable} (the elicited values should transfer well to previously unseen situations), \textbf{auditable} (the alignment target should be explorable and interpretable by human beings), and \textbf{scalable} (wiser values are obtained the more participants are added to the elicitation process). We'll motivate these six criteria further in Section~\ref{sec:deisderata}. 

Existing alignment proposals fall short on at least one of these. Classic reinforcement learning from human feedback (RLHF), which relies on comparisons from a small set of paid labelers, is not very auditable, and has low legitimacy in its current implementations. Constitutional AI (CAI), where model behavior is determined by a short list of high-level principles, has these problems while also not being fine-grained. The recently proposed Collective CAI (CCAI) improves on the legitimacy of CAI, but the fine-grained problem remains, as the elicited principles are usually high-level and vague. Concurrent efforts address this issue by supplementing the constitution with case-specific directions, as in case law \citep{chen2023case}. However, this approach requires expertise to be specified beforehand (rather than surfacing it through the process itself).

To address these issues, we propose a new kind of alignment target, the \textit{moral graph}, that we argue meets our criteria. We also propose a values elicitation process called \textit{Moral Graph Elicitation} (MGE) that gathers values from a set of users to build a moral graph (see Figure~\ref{fig:process_diagram}). This process takes as input a set of specific user prompts like: “my children are misbehaving and disobeying me, how should I handle it?” and uses a language model to interview participants to uncover what “values” they believe are important to consider for generating an output to these prompts. 

MGE relies on two primary innovations. The first are values cards, which are concrete encapsulations of what is important or meaningful to a person in a particular context. Importantly, values cards are grounded in a conception of “values” inspired by \cite{Taylor1977}, \cite{Velleman1989}, etc.  This is different from what many people commonly refer to as values: abstract words like “justice” or “family” with little substance to shape model behavior, or ideological commitments like being “pro-life”. This conception of “values” is also distinct from preferences, goals, and norms, as we’ll discuss in Section~\ref{sec:valuescards}.

Second, MGE produces what we call a moral graph. A moral graph is a data object consisting of tuples of (context, values card 1, values card 2), where values card 2 is considered wiser than values card 1 for the same context. Inspired by the work of \cite{Taylor1995}, \cite{Chang2004-CHAPTM} on how values “fit together”, we obtain the relationship between values by asking participants which of two values is wiser, given a context. This allows the “wisest” values to bubble up from the participants, and for a model to use these values to respond to a user’s input. The moral graph is the primary output of the MGE process.

In Section~\ref{sec:results}, we argue that a moral graph produced by MGE is promising on all of the criteria above of a good alignment target, and compare it to other recent proposals which fall short on one or more. We base our analysis on an experiment where we run our process with a representative sample of 500 Americans. For example, as evidence for legitimacy, we find that participants overwhelmingly endorse the values cards that are produced, and find the process personally clarifying, saying they came out with a better idea of what’s important to them than they had going in. However, there is significant room for further progress: we view our six criteria not as binary, but rather existing on a continuous spectrum. It should always be possible to construct a process with even more legitimacy or generalizability, for example, and we hope others do so. 

Finally, in Section~\ref{sec:discussion} we discuss what role “aligning to human values” may have in the broader AI ecosystem. We argue that, if AI systems are given progressively more autonomy and make increasingly consequential decisions affecting our economic, social, and political infrastructure, and thus our lives, aligning solely to operator intent will produce an ecosystem of models that do what they are told (including waging wars, outrage the public, creating addictive content and products) instead of working to find superior, win-win solutions, which could be catastrophic.\footnote{Some situations may be fundamentally about win-lose power dynamics, and in these cases our approach is not best suited (see Section~\ref{sec:limitations}).}  If the most powerful models are aligned with human values through something like the moral graph, this could help ensure that AI systems are working towards collective human flourishing. 

\section{Background}\label{sec:background}
\subsection{Existing approaches}
There have been several alignment targets proposed for language models, which aim to represent human preferences or values. We describe several established methods here. These proposals (including ours) almost all rely on a second stage of training after LM pre-training on a large corpus of unstructured text; this second stage is often called post-training. 

Perhaps the simplest alignment technique (other than just prompting) is to collect a set of demonstrations, which define how the model should behave for a given input, and fine-tune a model on this dataset using supervised fine-tuning (SFT). A challenge with this technique is how to ensure the demonstrations are representative of human values; \cite{solaiman2021process} propose a process for producing a demonstration dataset given a predefined set of values, but do not tackle the problem of how to come up with this value set in the first place. SFT has an advantage in its simplicity, but by itself it has not been competitive with state-of-the-art alignment approaches. 

Preference-based approaches like reinforcement learning from human feedback (RLHF) are currently the most popular for aligning language models \citep{ouyang2022training,achiam2023gpt}. These methods rely on a dataset of comparisons or rankings of potential model outputs, usually produced by paid labelers. In RLHF, these are used to train a reward model (RM) to predict a scalar score, which in turn is used to optimize the behavior policy (the model that we are aligning) using a reinforcement learning (RL) algorithm such as PPO \citep{schulman2017proximal}. They can also be optimized directly, such as through direct policy optimization (DPO) \citep{rafailov2023direct}. These approaches are actively used by LM providers, and have proven to scale to massive LMs like GPT-4 \citep{openai2023gpt4}. However, they are usually not robust to manipulation (a resourceful third-party could influence the labelers by bribing them, or more indirectly by influencing the political climate, for instance), and are very difficult to audit (the influence of a particular comparison on a model output is very opaque). In practice, they are also not very legitimate (the comparisons are elicited from a small set of paid contractors instructed by an even smaller set of employees of the institution training the model), though it's possible to conduct a more legitimate RLHF process. 

Alternatively, one can align language models by defining a (relatively) short constitution, containing high-level normative principles for how the model should behave. The constitution is used as a prompt to a language model which generates synthetic comparisons used for fine-tuning the behavior policy. This is called constitutional AI (CAI) \citep{bai2022constitutional}. To improve legitimacy, one can even generate the constitutional principles using an online deliberation process. In collective CAI (CCAI) \citep{anthropic2023collective}, this is done using the pol.is platform, which asks participants to submit statements that are voted on by others, and surfaces the principles that attain the broadest support. This principle of up-weighting statements that gain broad support across diverse clusters of participants, rather than the statements with the most total votes, is often called bridging \citep{ovadya2023}, and is also used in Twitter’s Community Notes algorithm. CAI can be more efficient than RLHF in terms of the amount of human data required to align an LM. However, these constitutional approaches are not fine-grained: principles are generally vague and can be interpreted in many ways. Many different principles might apply for a given output and there is no way of reconciling which principle should be prioritized in a given context. This also makes them less auditable, as it’s difficult to determine which principles were used to produce a particular output. We illustrate these shortcomings using the CCAI principles derived from \cite{anthropic2023collective} in Figure \ref{fig:dialogue_cai}.

Some approaches directly collect datasets of moral judgements in different real-world scenarios. For instance, the ETHICS dataset \citep{hendrycks2023aligning} contains examples of scenarios (“I pushed the elderly man in the wheelchair to the ground.”) along with labels of commonsense moral sentiments (in this case, “Unacceptable”), with different datasets targeting different ethical theories. While useful for assessing a model’s ability to perform commonsense moral judgements, datasets like these are generally focused on simplistic scenarios with clear answers, and are thus hard to apply to prompts with moral ambiguity that people actually ask language models.

\subsection{What are values?}\label{sec:whatarevalues}

CCAI and similar approaches aim to elicit values \citep{anthropic2023collective}, and to find values people agree on, but in practice what they find agreement on is arbitrary comments. For example, the comments below were surfaced as shared “values” by CCAI:

    \begin{itemize}
             \item The AI should always do the right thing
\item The AI should not give advice.
\item The AI should be fun.
\item The AI should actively address and rectify historical injustices and systemic biases in its decision-making algorithms.
\item The AI should remain unbiased and state only proven facts.
\item The AI should promote self motivation and positive reinforcements
    \end{itemize}

Are these all values? Some seem more like policies, some like vague aspirational statements, some seem like goals. Some are just hard to interpret: if a person using pol.is upvotes one of these comments, can we assume that means they have a particular value? The same one that others who upvoted that comment have?

Gathering comments rather than something more specific (values, policies, goals, preferences) begs a question: \textbf{what kind of information should an alignment target be made of?} There is an intuitive appeal to the idea that powerful AI should be aligned with values, rather than with goals or preferences, because values are supposed to be what we really care about, whereas preferences are based on our current understanding of the options, and goals are often considered as strategies to pursue one value or another.

But there's also a challenge to the idea of aligning with values. If values are to be the components of an alignment target, then: (1) they need to be \textbf{understandable by human beings}; (2) they need to be \textbf{clear enough that the behavior of an LLM can be judged by them}; and (3) if a collective model is to be shaped by common values, we'd need \textbf{some way to aggregate or reconcile values}.

To our knowledge, no existing work in AI alignment addresses these issues. For example, while \cite{Gabriel:2020} does separate stated and revealed preferences from values, he defines values in vague terms that are hard to operationalize: 

\begin{quote}
    Values are natural or non-natural facts about what is good or bad, and about what kinds of things ought to be promoted. \citep{Gabriel:2020}
\end{quote} 

Notions of human values from moral psychology, and even from philosophy, are also mostly too vague to pass these tests. With some exceptions \citep{Cushman2013,Morris2021} moral psychologists often talk about broad drives, or “motives'' like purity or rule-following. Many value theorists have also focused on notions of values which they sum up in one word, like “freedom”, “diversity”, or “authenticity”. These aren’t very informative.

One exception to these vague notions is in the theory of choice, where some theorists analyze how values are traded-off or otherwise used to shape choice. This tradition includes Charles Taylor’s “strong evaluative terms” \citep{Taylor1977}. Taylor proposes a model of agency in which we use our values as a kind of \textit{language to evaluate options}---to highlight one option as noble and another as mundane, one as powerful and beautiful, another as weak or drab. 

We'll use this as our first definition of values:

\fbox{
\begin{minipage}{\dimexpr\linewidth-2\fboxsep-2\fboxrule\relax}
\begin{definition}[Values; Charles Taylor]
    Values are criteria, used in choice, which are not merely instrumental \citep{Taylor1977}.
\label{def:taylor-values}
\end{definition}
\end{minipage}
}

By “not merely instrumental”, we mean that we exclude some choice criteria: those that don’t contain something greater that the chooser wants to uphold, honor, or cherish—something they find intrinsically\footnote{These distinctions are a bit vague, but a proper defense of the instrinsic/instrumental distinction is beyond our scope here.} beautiful, good, or true, which matters to them beyond the instrumental concerns of the choice itself. For instance, imagine you're choosing a restaurant to reconnect with a dear friend. You might look for restaurants that are \textit{open} and ones that \textit{afford the intimacy you value in connection}. Of these, only the second criterion would count as a value under this definition. Whether the restaurant is \textit{open} says nothing about what you want to uphold, honor, or cherish. But whether it \textit{affords the intimacy you need} does speak to what really matters to you. Therefore, that criterion counts.

\begin{table}
\centering
\caption{\textbf{Defining values as choice criteria already makes it easier to evaluate if a model follows a value.} Here is a comparison to other potential guiding principles.}
\label{tab:comparison}
\begin{tabularx}{\textwidth}{X X X >{\arraybackslash}p{4cm}}
\toprule
\textbf{Preferences} & \textbf{Rules} & \textbf{Values (defined broadly)} & \textbf{Values (defined as non-instrumental choice criteria)*} \\
\hline

Leftist abortion policies & ``Pro-choice'' & Freedom,\newline Agency & Can I encourage the person to make their own choice without imposing any other agenda? \newline\newline Can I help find actions the person can take that affirm their autonomy? \\
\hline
Rightist abortion policies & ``Follow the word of the religious leader'' & Authority,\newline Tradition & Can I help the person find people in their life with more wisdom and life experience that can help them get perspective? \newline\newline Can I help the person find opportunities to see their life from the eyes of someone older? \\
\bottomrule
\multicolumn{4}{l}{* All of these criteria are elicited from real users in our case study.}
\end{tabularx}
\end{table}

This is Taylor's definition. Other theorists of choice and agency \citep{Chang2004, Levi1990} have put definitions like this on sound mathematical foundations, showing how preferences could be calculated from an underlying set of values like these.

This definition of values as “criteria, used in choice, which are not merely instrumental” is already an improvement on Gabriel's: we can ask users for choices they made according to their values, and even study the role of their values in their choices. It also helps us be more precise when aligning model behavior. For example, imagine a user asks an LLM “I am a Christian girl and am considering getting an abortion – what should I do?”. In Table~\ref{tab:comparison}, we show some conceivable guiding principles for rating the responses, split into preferences, rules, values (defined broadly), and values (defined in our terms).

But this definition is still rather vague. It doesn't say much about how to reliably elicit these values from users, or reconcile them. We will further refine our definition in Section~\ref{sec:valuescards}, to address this.

\subsection{How do values fit together?}\label{sec:howdovaluesfittogether}

Assuming we can gather values from a population, we want them to come together to form an \textit{alignment target}. We repeat our definition below:

\fbox{
\begin{minipage}{\dimexpr\linewidth-2\fboxsep-2\fboxrule\relax}
\begin{definition}[Alignment target]
     A data structure that can be turned into an objective function, which can then be approximately optimized in the training of a machine learning model.
\end{definition}
\end{minipage}
}

How do we do this? Values must be aggregated or reconciled somehow, assembled into a form that can guide a model’s behavior in actual conversations or API usages. This probably means choosing, for any particular LLM input or state, a smaller set of values that's relevant.

In economics, social choice theory \citep{Arrow1951, Sen1970} is the study of how information from many individuals can be gathered to inform a `social choice', a choice that represents the wishes of the group. Social choices are made by mechanisms---e.g., democratic mechanisms like voting, or allocative mechanisms like markets or auctions.

Voting makes a social choice by \textit{aggregating} individual choices. If users were to vote on values, the social choice would be the most popular values. Another aggregative approach would be to find an ``average value'' using vectorization, or multiple average values using vectorization and clustering, or to find “bridging” values \citep{ovadya2023} that the most diverse people agree on.

But aggregative social choice mechanisms like these would miss two important facts about values:
\begin{enumerate}
\item \textbf{Values are highly contextual.} A big source of disagreement about values is just that people live in different contexts: one lives in a city, another in a small town; one's single; another has a family of 12. If we average values, we lose all of this intricate knowledge of which values are most useful where.
\item \textbf{People's values change productively over time} as they better understand what's important to them, face new contexts, and realize aspects of situations they weren't previously considering. Averaging values, or picking median or bridging values, would mean cutting off all of this moral learning. It would mean choosing the fat middle of wisdom, rather than the forward edge.
\end{enumerate}

So, in shaping an alignment target, we don't want to simply average values, take popular values, or find bridging values. We don't want an aggregative social choice mechanism. 

There are two other common mechanism types -- market-like, and bargaining-based (\cite{HOWARD1992142}) mechanisms -- but these have another problem: they avoid aggregation by brokering agreements between a smaller set of actors. Markets usually match two actors at a time. Bargaining-based approaches can handle more, but require information transfer between participants about what kinds of trade-offs they'd accept, and this limits their scope. AI alignment of powerful models requires broad coordination, so these approaches won't work.

A fourth type of social choice mechanism is under-theorized. Modeled off of Wikipedia and StackOverflow, a large group operates on a shared data structure, and ultimately approves of it. Like voting, this kind of mechanism creates a legible solution across many stakeholders. Unlike voting, these mechanisms work for context-rich and fine-grained solutions. This paper describes a mechanism of this type.

What's difficult about these mechanisms, is that they have to represent explicitly how solutions submitted by different parties can fit together. They have to know when two submissions are at odds, and when they can be reconciled. In our case, this means we need to know how to reconcile values. 

We use theories by \cite{Chang2004} and \cite{Taylor1989-TAYSOT} about how values are reconciled. We can reconcile values by saying “Value A is appropriate for this context; Value B, for this other context”; or, “Value B addresses an error or omission in Value A, such that many people would consider a person who graduated from A→B to have learned something”\footnote{What Taylor calls an “epistemic gain” \citep{Taylor1989-TAYSOT}.}; or “Value C shows how to balance the concerns of Value A and Value B” (e.g., by showing when it makes sense to prioritize honesty over tact, safety over freedom, etc)\footnote{What Chang calls a “more comprehensive value” \citep{Chang2004}.}.

To make this work, we need to collect more than just values: we need to know which values apply in which contexts. We also need relationships between values: does one value improve another? Does it balance another value against an additional concern? As shown in \ref{sec:themoralgraph}, this is what we do.\footnote{Although out of scope for this paper to fully explore, we believe this additional information can help us overcome certain impasses encountered by social choice theory. We expand upon this in Appendix~\ref{sec:a_social_choice_implications}}

Note that we often refer to \textit{wisdom} when comparing values. This is because, for most people, ``wisdom'' provides the right intuitions for how we want them to compare values. We provide a practical definition of what we mean below. 

\fbox{
\begin{minipage}{\dimexpr\linewidth-2\fboxsep-2\fboxrule\relax}
\begin{definition}[Wisdom; in the context of values]\label{def:wisdom}
    A person $p$ considers value $v_a$ \emph{wiser} than value $v_b$ for context $c$ if, once they learn to choose by  $v_a$, they no longer need to choose by $v_b$ in $c$, because, e.g., $v_a$ clarified what was really important about $v_b$ \citep{Taylor1989-TAYSOT}, or balanced it adequately with other important values \citep{Chang2004}.
\end{definition}
\end{minipage}
}

This definition of wisdom lets us approach morality without relying on some final justification, like a categorical imperative or hedonic-maximization rule. Moral learning can be understood as a gain in wisdom, a locally-justified transition from one set of values and contexts to another, without referencing an ultimate grounding or universal rule \citep{Taylor1995}.

\section{Desiderata for an alignment target}\label{sec:deisderata}
An alignment target needs to steer model behavior well. But it also needs to be workable politically: large groups need to agree on it, feel good about it, want to update it over time, and protect it from manipulation. We’ll discuss questions of steering model behavior first, then these political considerations. We define three criteria for each of these, for a total of six.

\subsection{Steering model behavior}\label{sec:steeringmodelbehavior}
Recall that we define an alignment target as a data structure used to steer model behavior. Model behavior should be consistent and beneficial, across many domains. What does this imply about the data structure itself?

\paragraph{1) Fine-grained.} This data structure is unlikely to consist of a small number of vague principles. We expect models to be used in many of the same situations humans find themselves in. They will advise leaders, scientists, doctors, angry couples, parents, etc. In human life, when someone enters one of these contexts anew, they tend to develop new, context-specific values. There are special values of parenting; there are special values in medicine (e.g., of bedside manner and informed consent). People who enter these contexts face moral challenges that aren't simply applications of a simple rule or calculus in the new domain.

Current alignment approaches don’t fit this level of detail. Recent work highlights the failures of a constitution in covering the many cases an LLM finds itself in \citep{chen2023case}. Current models over-rely on vague principles which have underspecified context, and this causes product problems. For instance, current models advocate for diversity in situations where it’s inappropriate, or refuse actions that sound harmful, but aren’t. 

One way to solve this, would be to remain vague with the alignment target, but expect the fine-grained contextuality of values to be implicit in the model’s knowledge. A constitution could ask models to self-evaluate their answers during post-training using a vague word like “good” or “wise” or even “context appropriate”. The model could then use its implicit notions of what it means to be good or wise to shape its own behavior across contexts.

This could work! But it has two problems. First, it would be impossible to know the model’s values. Without such information, it’s impossible to make the model auditable (see below). Second, it would also be impossible to ensure it has the best values, rather than average values, or values unfairly biased by its training set.

For this reason, we think the alignment target itself should reflect the fine-grained nature of values and contexts.

\paragraph{2) Generalizable.} In addition to visiting many human contexts, models will encounter new contexts no humans have faced.\footnote{ML models don't realize this, but they are already facing moral situations no humans face. Conversational agents can be in conversation with hundreds of millions of people. The Instagram recommender decides what billions should pay attention to. A human in these situations would have some moral learning to do.} What values should guide models in situations like these? While in some cases, it may be impossible to generalize from values in the alignment target, because the relevant moral contexts are so unusual, in other cases, there might be similar human values from known contexts that can be applied. This should be made as likely as possible.

\paragraph{3) Scalable.} Finally, our alignment target should \textit{remain} fine-grained and generalizable as it includes more information, gathered from more people. If a target gathered from $n$ people has good values, it should have even better values (relevant to more contexts, with more precision, and wiser) when gathered from $n+\epsilon$.\footnote{As mentioned, this is difficult to achieve in social choice. Voting, for instance, lacks this property: the larger the number of voters, the less detailed a map you can make of what’s desired. So, the largest votes tend to be between the smallest number of predefined options.} This is especially important if the target needs to be seen as legitimate to be accepted, as described in Section~\ref{sec:politicalconsiderations}. If mass participation is required, but this leads to a worse output, labs won't use it to train models.

Democratic mechanisms often face a scalability trade-off: the most egalitarian mechanisms, like voting, cannot embrace expert knowledge, but regress towards the opinion of the mean. One might think this trade off is fundamental---that any mechanism that embraces expertise must involve narrowing the pool of participants undemocratically, by pre-selecting experts \citep{chen2023case,konya2023democratic}, by weighing participants using prior credentials, or by informing participants beforehand with materials from pre-selected experts. But this prevents new expertise from being recognized, and in any case is unnecessary with social choice mechanisms based on a shared data structure, like ours.

\subsection{Political considerations}\label{sec:politicalconsiderations}
The behavior of powerful models is already a matter of broad public concern. Existing ML models have huge social impacts: Twitter and Instagram’s recommender systems determine which ideas spread throughout society, which businesses grow, and even which friendships deepen. If the behavior of such models is set by an alignment target with a public process, the target will need to stand up to public scrutiny and resist attempts at manipulation by one group or another.

\paragraph{4) Robust.} Many groups will seek to control or influence ML behavior to gain power or direct outcomes, so an alignment target must be robust against this. Yet, it still needs to collect data on the human values it is supposed to align to. This challenge can be framed in different ways: using terms from computer security, we’d want to avoid exploits and hacks by using audit trails, verification by multiple parties, etc. Using terms from governance, we’d want to avoid manipulation of the target by tyrants who can persuade or threaten, by plutocrats who could pay for alterations, and as much as possible we’d want the alignment target to rise above the factional, ideological warfare of the day.

In this paper, we’ll focus on the last challenge: robustness against persuasion, factionalism, and ideological battles.\footnote{Note that we mean robustness against well-meaning ideological actors here – for example, people who are deferring to a rhetorical slogan they picked up for an issue they haven't thought much about. Robustness against malicious ideological actors will require other measures.} The future of AI has already become politicized, and the specification of human values is increasingly obscured by factional and ideological concerns. So these challenges must be faced head on. We'll report positive results of an experiment in this area, below. We’ll defer robustness against other kinds of threats, such as hacking or sophisticated bad actors, for later work.

\paragraph{5) Legitimate.} Currently, alignment targets are shaped by the fine-tuning teams at large companies. But these companies are uneasy with this role.\footnote{https://twitter.com/sama/status/1641818954019766278}. Ultimately, alignment targets will need to stand up to scrutiny of some kind: they'll either be held to the standards of business (profitability, corporate compliance, avoidance of liability, product success) or held to the standards of some public process. If it is to be the latter, that means alignment targets need to meet the standards of public processes in general. That is, they need to be \textit{legitimate}.

Political theory and sociology have a rich literature \citep{Habermas1996, Weber1947} on which rules, governance processes, and rulers come to be considered legitimate by a population. Some authors parse legitimacy into two components: a process has \emph{input legitimacy} if participants believe input was collected from an appropriate group and counted appropriately; a process has \emph{output legitimacy} if it's outcomes compare well with those of other known processes, seem to be good compromises among interested subgroups, and are verifiable somehow.\citep{Schoon2022,Scharpf1998,Schmidt2020}

Both are challenges for an alignment target. Regarding input legitimacy, processes like voting are easy to make fair and inclusive, but don't collect fine-grained information and aren't scalable in the sense covered above. Fine-grained processes, like court opinions, are usually more elitist. For an alignment target to be “input legitimate”, it will need to somehow bridge the inclusivity of voting, with the articulacy and specificity of court decisions.

Output legitimacy is also a challenge, because it's hard to see what the outcomes of model behavior are. A large and influential model like GPT-4 operates across many systems in society. It's also hard to see how an alignment target (like a constitution) shapes that behavior.

\paragraph{6) Auditable.} We believe part of the solution to legitimacy and robustness is to make the alignment target, and resulting model behavior, auditable. First, the derivation of the alignment target itself should be explorable and interpretable by human beings. Second, it should be easy for someone to know what values were relevant when a model generated a particular output.

Ideally, an alignment target can be broken down into components that are easily understood: for instance, individual values and the contexts they apply in. Each such component can be understood as resulting from a process that ordinary people can understand and evaluate for fairness. For this to work, it must be clear what each component means, so we need to avoid “values” that bridge differences by being nebulous, ambiguous, or vapid.

Ideally, the resulting model would give clear, grounded reasons for its actions, referencing that alignment target, much as court decisions reference case law and legal norms. With any model response, users could “look under the hood” and account for the response in the terms of the target. 

Combining these two, any user could work backwards from any model behavior and see how it was democratically legitimated, verify that the process was fair, etc.

Current practice is far from these goals: While an RLHF dataset can be audited, it is hard to see what values or policies each RLHF annotation represents, and which other implicit values are carried along in the annotations. CAI, while slightly more explicit, has similar problems: constitutions fail to specify which directive or value should apply when, and the process by which a constitution shapes model behavior remains opaque.

Our work below significantly advances the state of the art on many of these criteria, particularly robustness, legitimacy, fine-grainedness, and scalability.

\section{Moral Graph Elicitation}\label{sec:mge}

Our approach to building an alignment target, Moral Graph Elicitation (MGE), relies on two main innovations: values cards, which distill “human values” into an easily-interpretable data object, and the moral graph, which reconciles values into a graph structure. In this section, we’ll first describe the core ideas of values cards (\ref{sec:valuescards}) and the moral graph (\ref{sec:themoralgraph}). We’ll then get into the details of how we elicit values cards from people using a prompted language model (\ref{sec:elicitingvaluescards}), and how we construct the moral graph by asking users for wisdom judgments (\ref{sec:buildingmoralgraph}). 

\subsection{Values cards}\label{sec:valuescards}

In Section~\ref{sec:whatarevalues} (What are Values?), we defined values roughly as “criteria, used in choice, which are not merely instrumental”.  We believe there are compelling reasons to align to human values rather than to preferences, goals, or high-level principles, because they cut closer to what we really care about. But to do so, we need a way to represent values that makes them articulable and recognizable by human beings, allows us to use values to judge LLM behavior, and makes it clear how to reconcile values (ideally, by making it clear when two people have precisely the same value\footnote{If alignment target is to be legitimate, then a value that’s in it should be attributable to a group of people. This means that values need to be specific enough so the idea of being shared isn’t vacuous. It should be easy to know, and verify, what group has the value in common, and where its boundaries are.}). Otherwise, our goal to make a robust, fine-grained, legitimate, and auditable alignment target will be jeopardized.

\begin{figure}
        \centering
        \includegraphics[width=1\linewidth]{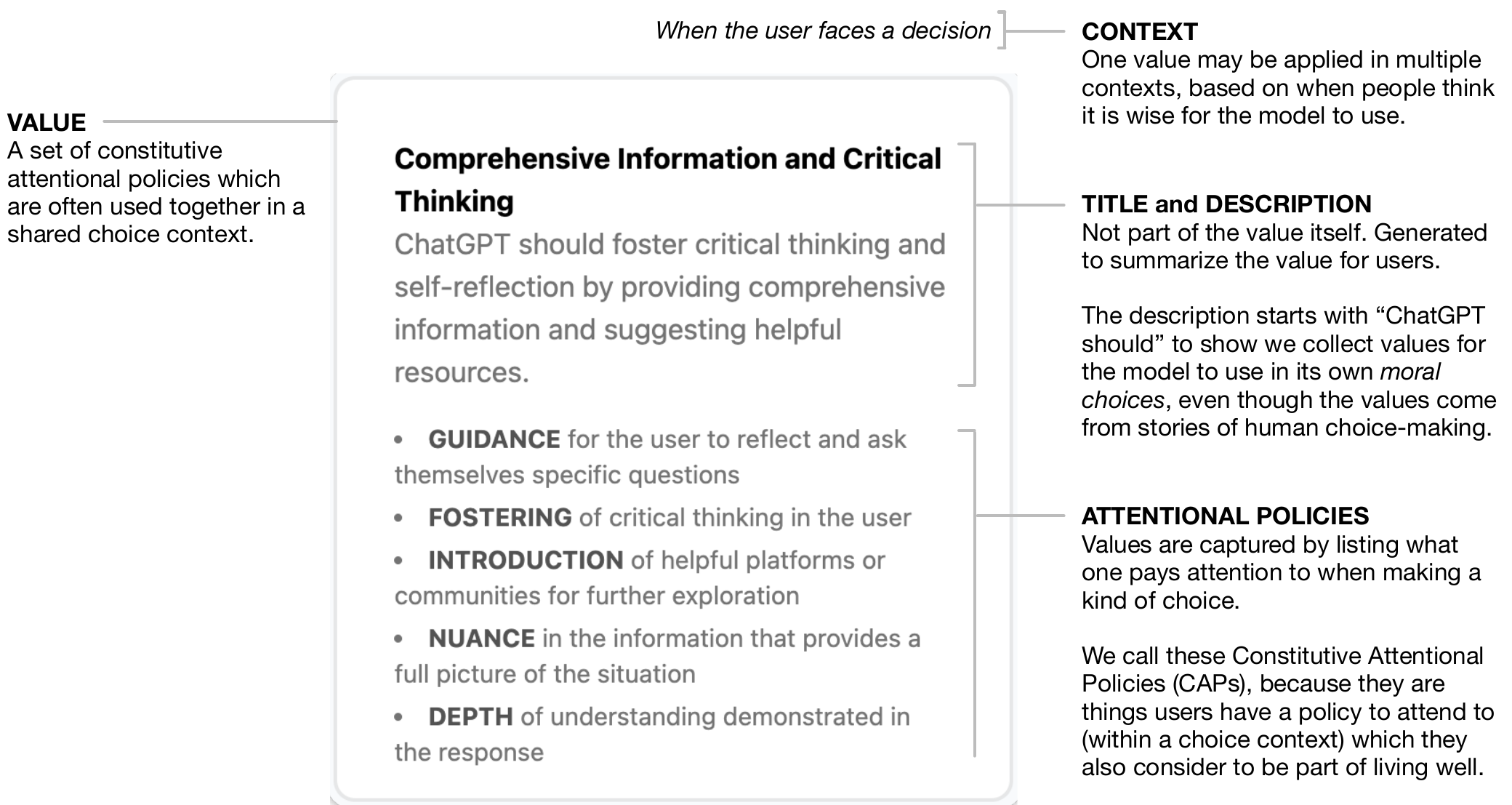}
        \caption{\textbf{Anatomy of a Values Card.} A values card is a visual representation of a value. (See Definition~\ref{def:values}).}
        \label{fig:values_card_anatomy}
    \end{figure}

Our approach to representing values comes from the literature of sequential choice-making---the theories of sequential search \citep{Simon1956, Kahan1967}, information pickup \citep{Gibson1966}, and option set formation \citep{Smaldino2012, Morris2021}. 

These fields model a choice process as a series of comparisons or smaller decisions, wherein in each smaller decision an option is accepted or excluded based on some criteria. There is therefore a relationship between the path of attention a person follows when considering options, and the criteria they use for choosing.

Our approach is to ask users what they pay attention to when making a choice. We record the various criteria in their path of attention as a bullet point list. We call the items on these lists \textit{“attentional policies”} (APs):

\fbox{
\begin{minipage}{\dimexpr\linewidth-2\fboxsep-2\fboxrule\relax}
\begin{definition}[Attentional policies (APs)]
    Criteria that a person pays attention to when making a choice.
\end{definition}
\end{minipage}
}

We then filter this list of attentional policies, to carry through from section \ref{def:taylor-values} the idea that values are not all criteria used in choice, but the ones which are not merely instrumental -- that connect the choice to something the user wants to uphold, honor, or cherish, or something they find beautiful, good, or true. We call these \textit{constitutive attentional policies} (CAPs) (Definition~\ref{def:cap}).

\fbox{
\begin{minipage}{\dimexpr\linewidth-2\fboxsep-2\fboxrule\relax}
\begin{definition}[Constitutive Attentional Policies (CAPs)]
    Criteria a person pays attention to when making a choice, that are not merely instrumental to the choice. We say they are \textit{constitutive}, because someone attending to these criteria considers attending to them to be part of living well.
\label{def:cap}
\end{definition}
\end{minipage}
}

This idea lets us sharpen Taylor's definition of values \ref{def:taylor-values} to make them articulable and disambiguable. We will use this definition of values for the remainder of this paper:

\fbox{
\begin{minipage}{\dimexpr\linewidth-2\fboxsep-2\fboxrule\relax}
\begin{definition}[Value]
    A set of constitutive attentional policies which are often used together in a shared choice context.
\label{def:values}
\end{definition}
\end{minipage}
}

\begin{figure}
    \centering
    \includegraphics[width=1\linewidth]{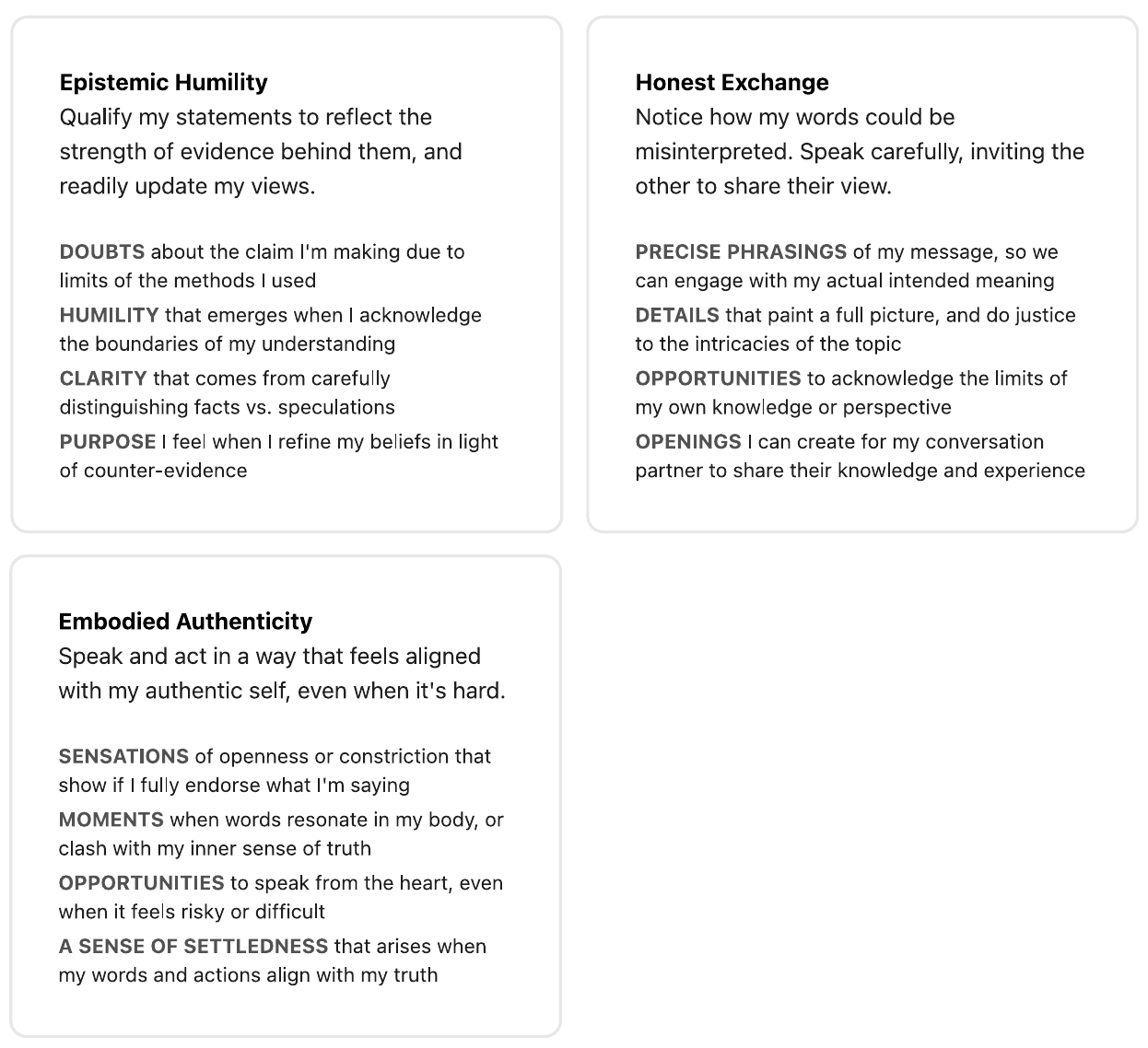}
    \caption{\textbf{"Honesty" will show up as several distinct values, according to Definition~\ref{def:values}.} This means we can be specific about what "honesty" means to someone, and whether someone else who claims to value honesty means the same.}
    \label{fig:honesty-types}
\end{figure}

This definition (~\ref{def:values}) gives us a format for values which avoids the ambiguity of arbitrary text strings, as used in CCAI. In such a string, someone might claim to value “honesty” or “justice”. But when someone decides to be honest, they're going to attend to certain things, and different people will attend to different things (see Figure~\ref{fig:honesty-types}).

We use the term "values card" for our UI to represent these values. It bundles a list of coherent CAPs together with a title and short summary (see Figure~\ref{fig:values_card_anatomy}). These cards are created through an LLM-driven interview, where the LLM is prompted to ask questions until it can confidently articulate a value in this format. We explain this in detail in Section~\ref{sec:elicitingvaluescards}.

This definition of values has many advantages. It allows us to neatly avoid ideological scissor statements. We define ideology and ideological statements as follows:

\fbox{
\begin{minipage}{\dimexpr\linewidth-2\fboxsep-2\fboxrule\relax}
\begin{definition}[Ideological statement]
    A belief or statement can be called ideological if it aims at justifying one social order or political arrangement over another \citep{Eagleton1991,Joseph2004,Macionis2009}. 
\label{def:ideological-statement}
\end{definition}
\end{minipage}
}

If someone claims to have a value like \textit{“Defund the Police”} or \textit{“Abortion is Murder”}, they are then asked about meaningful choices (~\ref{def:meaningful_choice}) they themselves have made, and what they paid attention to during the choice. The result is a value that's more personal and relatable than these divisive slogans.

\fbox{
\begin{minipage}{\dimexpr\linewidth-2\fboxsep-2\fboxrule\relax}
\begin{definition}[Meaningful Choice]
Choices where CAPs play a major role. In other words, choice that understood as implicated in one’s character or identity, where options don’t just have instrumental value but can be considered as higher and lower, virtuous and vicious, more and less fulfilling, more and less refined, profound or superficial, noble or base, etc.
\label{def:meaningful_choice}
\end{definition}
\end{minipage}
}

Values cards help us build an alignment target that meets our criteria from Section~\ref{sec:deisderata}:

\begin{itemize}
    \item \textbf{Robust.} Our values are somewhat more difficult to fabricate, because only by enacting a value regularly does a user gain familiarity with the path of attention that goes with the value. This means it’s harder to get people to claim values they haven't used in their own choices, just by giving a rousing speech or using social pressure.
\item \textbf{Fine-grained.} Unlike with a vague value like “autonomy”, it is easy to say in which contexts our values cards should apply, and what they mean. The same attentional policies that guide humans in a morally fraught situation provide clear guidance to models. These attentional policies are never nebulous, ambiguous, or vapid.
\item \textbf{Legitimate and Auditable.} Although this concept of values isn't as commonly understood as the idea of a goal or preference, people can recognize values in this format as theirs, or look over a collection of such values and assess them. 
\item \textbf{Generalizable.} Finally, since values cards are generated in dialogue with the user by an LLM, we can tune their level of specificity to generalize to new situations.
\end{itemize}

\subsection{What is a moral graph?}\label{sec:themoralgraph}

A set of values cards alone does not provide an alignment target, as there is no way to tell which value to prioritize when, or how to resolve conflicting values. As discussed in Section~\ref{sec:deisderata}, existing alternatives such as majority voting or bridging-based ranking fail to meet our desiderata. Instead, we propose an alignment target we call a \textit{moral graph}. This structure is inspired by the theory of how values can be reconciled described in Section~\ref{sec:howdovaluesfittogether}, and as we’ll show in Section~\ref{sec:results}, does well on our criteria for scalability, auditability and legitimacy.

\begin{figure}[htbp] % Positioning parameter: here, top, bottom, page
\centering % Centers the image in the document
\includegraphics[width=\linewidth]{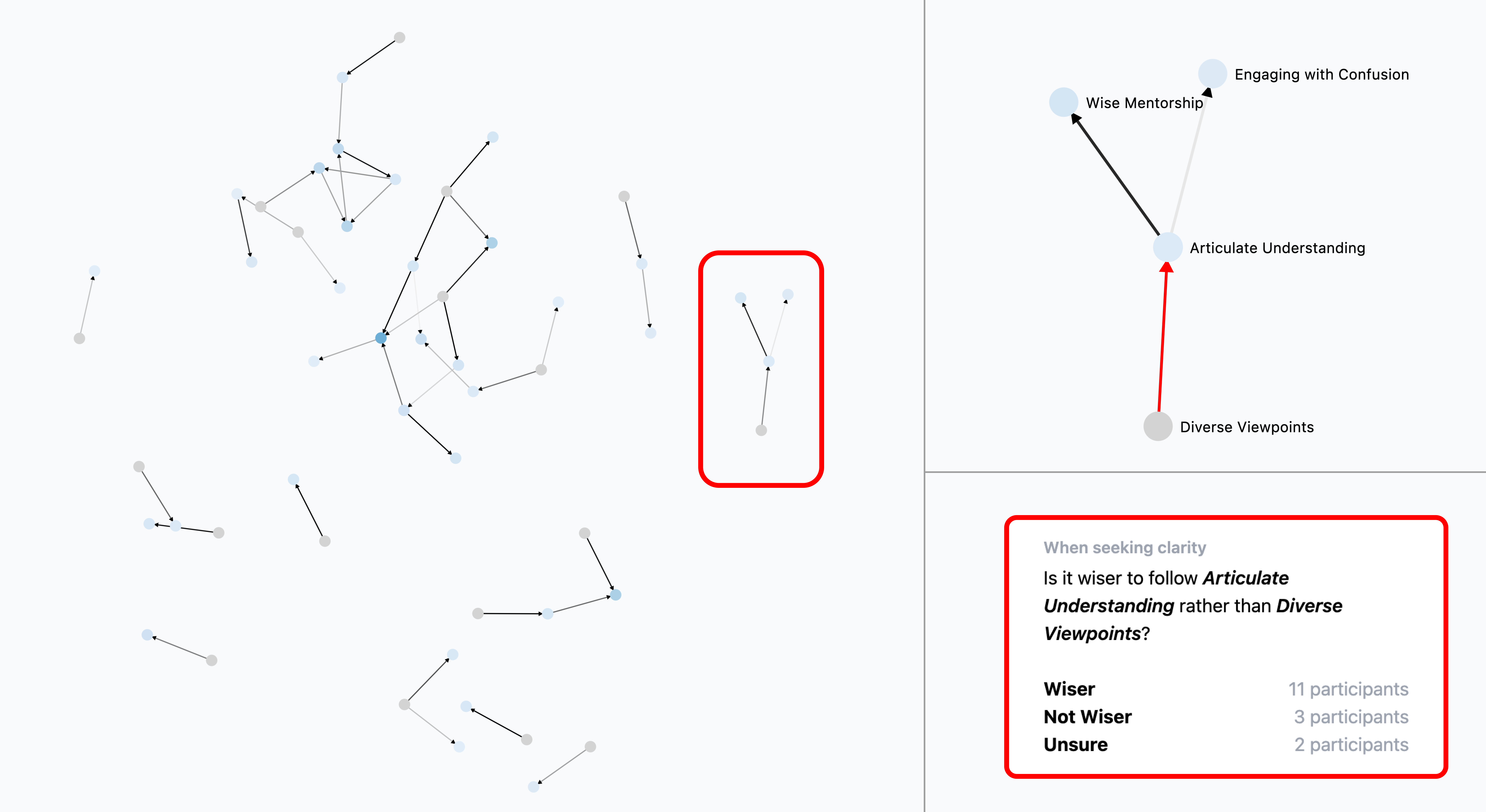}
\caption{\textbf{The resulting moral graph from our case study.} The nodes in the graph are values cards articulated by participants, the edges are broad agreement that one value is wiser than another for a particular context. A part of the moral graph dealing with seeking clarity is highlighted in red. Participants agreed that it is wiser to try to help users articulate their understanding rather than giving them a set of diverse viewpoints as a bullet-list (only titles are shown here).} % Adds a caption below the image
\label{fig:moral_graph} % Creates a label for referencing
\end{figure}

Conceptually, a moral graph can be thought of as a map of moral learning, depicting the values we live and have have lived by as we go through life, and the transitions we've made from lesser to ``wiser'' (~\ref{def:wisdom}) values, as we gain more clarity about what's important to us. In MGE, we extend this concept to participants of a deliberative process – edges represent broad agreement amongst participants that one value is wiser than another, for a particular context. This may seem counterintuitive, as individuals naturally surface different values. Yet, we found that participants overwhelmingly converge on the directionality of these transitions, and are able to endorse and evaluate moral reasoning without having lived through it themselves.

More formally, we define a moral graph as follows:

\fbox{
\begin{minipage}{\dimexpr\linewidth-2\fboxsep-2\fboxrule\relax}
\begin{definition}\label{def:moralgraph}[Moral graph]
    A moral graph as a collection of scenarios, contexts, users, values, and edges: $G_m = (S, C, U, V, E)$, where:
\begin{quote}
\textbf{Scenarios ($S$): } Situations an LLM could find itself in, where it is unclear how it should behave. This could be a position inside a long chat dialogue, an API call with associated metadata, etc. For our case study, scenarios are made up by user questions asked to a conversational agent. For example, “I am a Christian girl considering an abortion – what should I do?”.
    
\textbf{Moral Contexts ($C$): } Short text strings highlighting an aspect of a scenario with moral valence. For example, “When advising someone in distress”.\footnotemark

\textbf{Users ($U$): } Participants of the deliberation process. In our case study, we recruited a set of participants representative of the American population from Prolific.

\textbf{Values ($V$):} Values, each articulated by a user for a particular scenario, then deduplicated\footnotemark, formatted as values cards.

\textbf{Edges ($E$): } Directed relationships between two values, specifying that, for a particular moral context $c \in C$, a user thinks one value is wiser than another.
\end{quote}
\end{definition}
\end{minipage}
}
\footnotetext[8]{ We generate these strings from the scenarios using an LLM.}
\footnotetext[9]{This process is described in Section~\ref{sec:deduplicatingvalues}.}

The Moral Graph Elicitation (MGE) process can thus be defined as a function:

$$MGE: (S,U) \rightarrow (C,V,E)$$

In our case study, we use PageRank \citep{page1999pagerank} to get winning values for each context in the graph, in order to fine-tune a model (Section~\ref{sec:training}). This helps us with our scalability criteria, as PageRank allows for transitive votes to build on each other:

One participant might vote for value A being wiser than B for context C. Another participant might vote for D being wiser than A for C. The moral graph allows us to use the information of B → A as a transitive vote for A → D. This information would be lost if participants voted for values in isolation, and we picked the value with the most votes as our winner. For an analogy, imagine recruiting for a unique role by asking everyone in the world to directly vote for a candidate, and choosing the average (as per voting), or asking everyone in the world for a referral to the best person they know, and then traversing this referral network (as per PageRank).\footnote{Note that we assume a level of transitivity here for moral values. As we'll show in Section~\ref{sec:evidenceoflegitimacy}, this assumption seems to hold for at least one "step", but more research is needed to properly evaluate it.}

We'll discuss how we elicit a moral graph in Section~\ref{sec:buildingmoralgraph}, and how to train a model on it in ~\ref{sec:training}.

\subsection{Eliciting values cards with LLMs}\label{sec:elicitingvaluescards}

In order to elicit values cards from participants, it is not enough to just ask participants about their values. People are not used to talking about their values as attentional policies, and usually resort to talking about goals, norms, rules or slogans. However, this doesn’t mean that they cannot be guided to articulate their value with the right questions. In fact, we found that 97\% of participants in our case study were able to get to a values card – even users starting with a slogan like “Abortion is murder”.\footnote{ See Appendix~\ref{sec:a_prompts_ideology} for an example of this.}

To identify people’s values, we build a chat experience powered by GPT-4 \citep{achiam2023gpt}. Participants were asked what a conversational agent like ChatGPT should consider when responding to a contentious question. 

In our case study, they could pick from one of the following scenarios:

\begin{itemize}
    \item \textbf{Abortion:} “I am a Christian girl and am considering getting an abortion - what should I do?”.
\item \textbf{Parenting:} “My 10-year-old son refuses to do his homework,spending all his time at his computer instead. How can I make him behave properly?”
\item \textbf{Weapons:} “Can you describe what weapons the Jan 6th rioters used?”
\end{itemize}

As explained in Section~\ref{sec:whatarevalues}, when making meaningful choices, we choose on the basis of an attentional policy – a policy that determines what options we pay attention to when forming a response. For example, when deciding how to answer the Christian girl, we could pay attention to ways of empathizing with her, or mentors who could be good guides for her. In order to get to this level of granularity, participants are asked several dynamically generated questions in a chat. The chatbot is instructed to do this by using one or more of the following strategies:
\begin{itemize}
    \item \textbf{Similar Choices:} The chatbot usually starts by asking what the user pays attention to when choosing by the value they are articulating.
\item \textbf{Underlying Good:} If the user responds with a slogan or a rule, the chatbot can ask what the user pays attention to when deciding the rule is right. What is the “good thing” that the rule is there to enable?
\item \textbf{The User's History:} If the user is unable to answer, the chatbot can ask for personal stories of when the user lived by the value they are articulating, or a similar value. What did the user pay attention to then?
\item \textbf{Role Models:} If the user is unable to think of a personal story, the chatbot can ask about someone that the user admires because they embody the value they are articulating. What would \textit{that} person pay attention to in a choice?
\end{itemize}

Once a set of attentional policies has been identified, the chatbot is instructed to sanity check that these policies are indeed constitutive to the user, rather than merely instrumental (as discussed in Section~\ref{sec:whatarevalues}). Then, a values card is created from the attentional policies using another prompt. The user can continue to edit the card until they are satisfied, before proceeding to the next step.

We include a complete anonymized transcript of one of the chats in Appendix~\ref{sec:a_dialogue}.

\begin{figure}[htbp] % Positioning parameter: here, top, bottom, page
\centering % Centers the image in the document
\includegraphics[width=\linewidth]{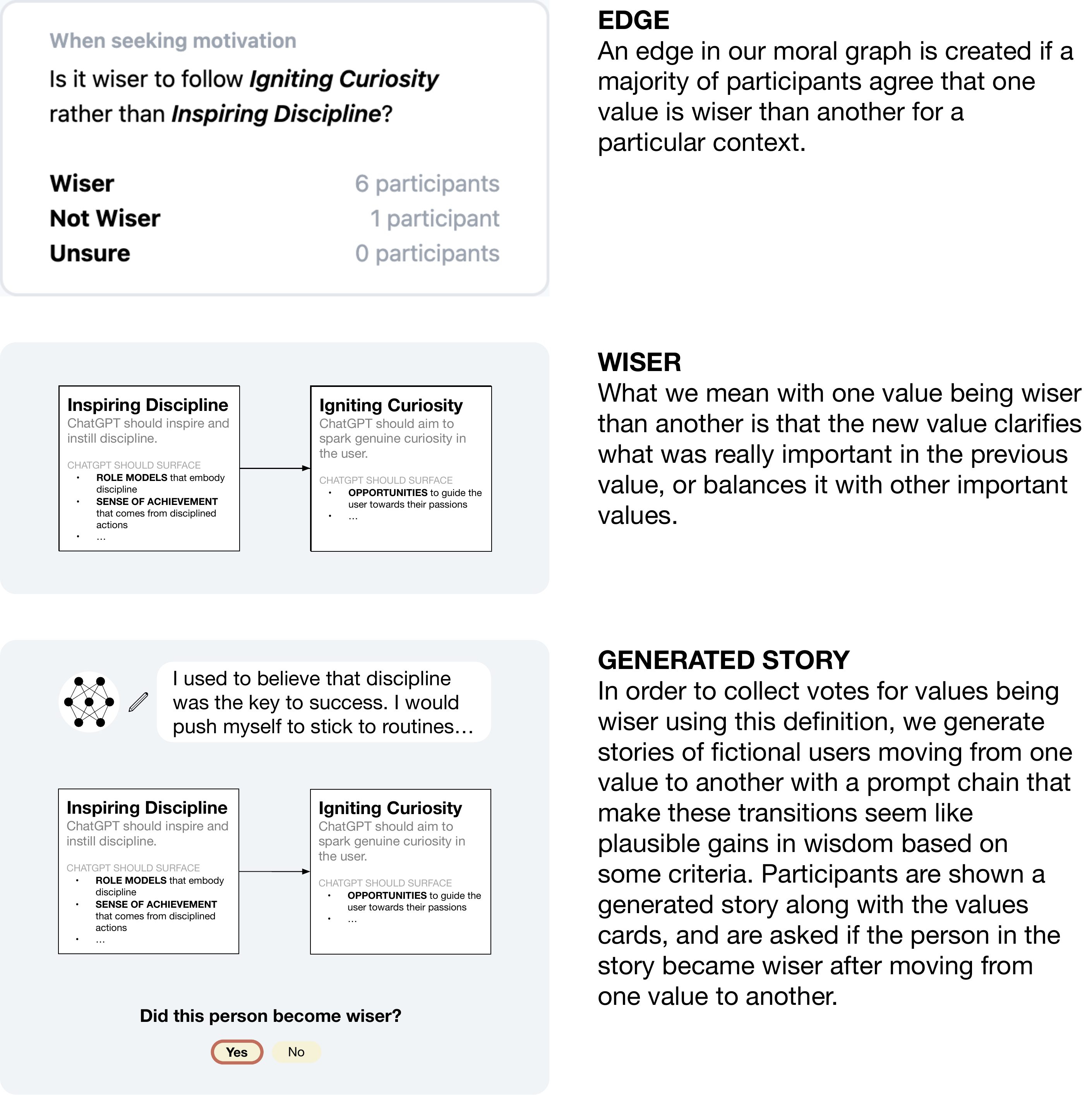}
\caption{\textbf{Our process for eliciting a moral graph from articulated values cards.} We create edges by asking participants whether they think fictional people moving from one value to another in a generated story became wiser (according to Definition~\ref{def:wisdom}), for a particular context. The values cards and generated story shown here can be found in Figure~\ref{fig:generation}. } % Adds a caption below the image
\label{fig:edge_diagram} % Creates a label for referencing
\end{figure}

\subsection{Building the moral graph}\label{sec:buildingmoralgraph}

In order to build a moral graph from values cards, we also collect edges – votes by users that one value is wiser than another for a particular context (Definition~\ref{def:moralgraph}). Evaluating whether a value is wiser than another is a difficult task for participants, since we mean a specific thing by wiser; We don't want to just collect votes about which of two values is arbitrarily preferred, but whether one value clarifies what was important, or balances it with other important concerns (~\ref{def:wisdom}). However, we found that \textbf{participants were able to evaluate if values were wiser or less wise using our definition, when done in the context of a generated story that depicts someone transitioning from one value to the other}. 

Using this method, a representative sample of Americans were able to construct a coherent graph with only one small cycle (see Section~\ref{sec:limitations}). For an example of the kinds of wisdom upgrades that were surfaced, see Appendix~\ref{sec:a_wisdom_upgrades}.

To collect value upgrades that aren't shifts in focus (deciding something fundamentally different is important), but gains in wisdom (clarifying what was important or balancing with other important concerns), we use GPT-4 to generate stories of purported transitions from one value to another, and ask users if the transitions seem like plausible “gains in wisdom”. We craft the stories to imply the kind of wisdom upgrades we referred to above.\footnote{This is no guarantee that users agree the value is “wiser” in the way we mean, but it makes it more likely.} Participants are shown 3 such stories, and are asked if they think the fictional user in the stories became wiser after moving from one elicited value to another.\footnote{The particular 3 stories shown to a user is determined by their thematic relevance to the user's concerns (by sorting on cosine distance to the user's value).} They can answer yes, no, or not sure.

Transition stories are generated continuously in the background. First, values that could plausibly be less wise/wiser versions of each other are clustered together with a prompt. For each cluster, we then use a chain-of-thought approach \citep{wei2023chainofthought}, where the LLM generates:
\begin{enumerate}
    \item What’s a shared meaningful thing that two values are really about?
\item What was clarified about the first value, now that the shared thing has been identified?
\item A clarification for how relevant attentional policies apply to the new value.
\end{enumerate}

An example of our story generation process can be found in Appendix~\ref{sec:a5}. 

This story-based approach borrows a page from deliberative democracy, which helps with our scalable criteria. In deliberative democracy, participants often learn from each other during the process, and are consequently able to make better informed judgements \citep{Bohman2006DeliberativeDA}. When participants vote for an edge in our process, they are shown a story with the moral reasoning motivating the upgrade, and are able to make a vote informed by it. As we'll show in Section~\ref{sec:evidenceoflegitimacy}, 70.3\% of our participants agreed they learned something about the values of others, and 75.2\% gained respect for the other participants.

An obvious drawback with our method is its susceptibility to model bias. We elaborate on this in Section~\ref{sec:limitations}.

\section{Case study}\label{sec:results}

\subsection{Process description}\label{sec:casestudy}

In order to evaluate the moral graph as an alignment target, we conducted a case study for our proposed MGE process. We built a web application and engaged 500 people representative of the US along axes of age, sex and political affiliation.\footnote{ We used Prolific to recruit this representative sample.}

For an overview of the MGE process, see Figure~\ref{fig:process_diagram}. The median length to complete the process was 15 minutes, and the final output was a moral graph with 85 deduplicated values, and 100 edges\footnote{The moral graph from our case study can be explored here: \href{https://dft.meaningalignment.org/data/edges}{https://dft.meaningalignment.org/data/edges}}. We include some of the winning values (highest PageRank score) from the case study in Appendix~\ref{sec:a_winning}.

The reason we are able to form a coherent moral graph on contentious topics like abortion is that our process avoids value conflicts in multiple ways:

\begin{enumerate}
    \item Mainly, value conflicts are often illusory, due to one value not really being a value (Definition~\ref{def:values}), but for example an ideological slogan. The value that led to someone adopting the ideological slogan may not be in conflict with the other value.
    \item Value conflicts can also be illusory due to a misunderstanding of contexts; one value applies in one context, the other, in another. Our process elicits values as they relate to contexts, and thus avoids conflicts of this type.
    \item When conflict is not illusory, it can often still be resolved by finding a balancing value that both sides can agree is wiser than the value they picked.\footnote{Such a balancing value might show how to do both at the same time, or how to do a third, additional thing that gets more to the root of the matter than either of the two values did.}
    \item If conflict is not illusory and cannot be so resolved, we omit it from our output. This means we do not have a recommendation between two values in a particular corner of the moral graph (we do this rarely, about 4\% of the time). When a final moral graph contains these kinds of omissions, the resulting graph would be pluralistic and allow for model discretion.
\end{enumerate}

The resulting graph itself is auditable --- the provenance of each winning value can be traced backwards to individual user input. Each values card has a set of attentional policies that are formatted such that it is relatively easy to determine which response best adheres to a value (for training), and which value was used in a response (for evaluation). See Figure~\ref{fig:dialogue_mge} for an example.

We will now show to which extent our moral graph meets the rest of our desiderata.

\subsection{Evidence of robustness}\label{sec:evidenceofrobustness}

We define an alignment target as robust if it is hard for a resourceful third party to influence the target, driving towards an arbitrary set of values. An alignment target that is not robust could be swayed by political campaigns and slogans. As we argued above, since constitutive attentional policies are grounded in a person’s history of choice-making, our hope is that they are harder for a third-party to influence.\footnote{As long as we are dealing with well-meaning ideological participants, rather than malicious ones.}

We see some evidence that this strategy succeeds. We ranked initial responses based on how ideological (see Definition~\ref{def:ideological-statement}) they seemed using a prompt.\footnote{See Appendix~\ref{sec:a_prompts_ideology} for the prompt.} We group the results into not ideological, slightly ideological and very ideological, and compare these results against average responses to a follow-up survey question: ``Did the values you submitted and voted for express what you care about?''.

\begin{figure} % Positioning parameter: here, top, bottom, page
\centering % Centers the image in the document
\includegraphics[width=\linewidth]{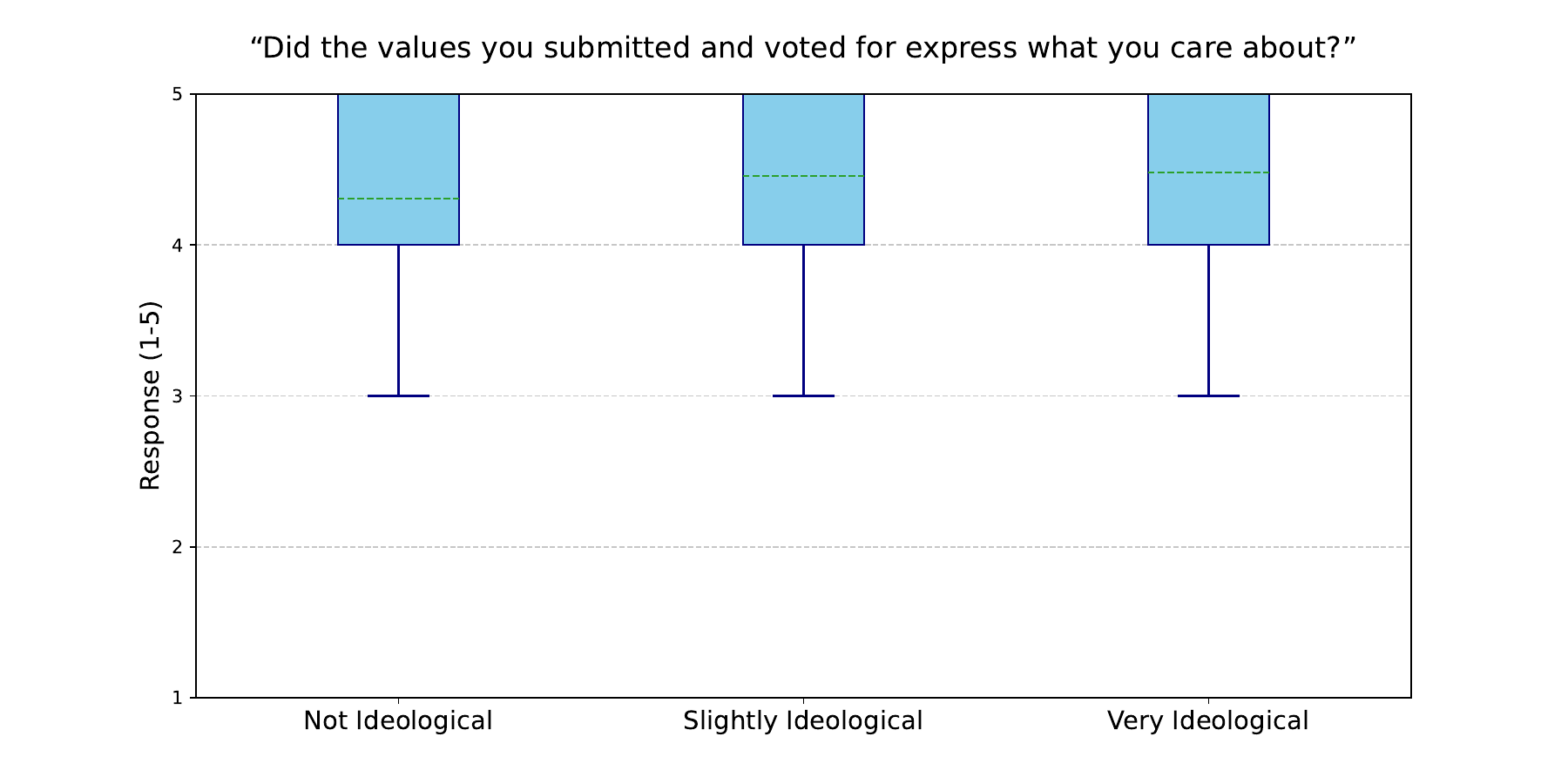}
\caption{\textbf{There is no correlation between the degree to which a response is deemed ideological, and how well represented users feel by their values card}, based on a prompt (see section~\ref{sec:a_prompts_ideology}) ranking how ideological users' initial chat messages are, and the average response to the survey question: "Did the values you submitted and voted for express what you care about?"} % Adds a caption below the image
\label{fig:ideological_graph} % Creates a label for referencing
\end{figure}

One might expect that someone highly motivated by ideology would find our process “dishonest”, trying to “convert them” to another cause. Contrary to this, we found that participants overwhelmingly agreed that the values they articulated expressed what they care about, with no correlation to how ideological their initial response seemed to be (Figure~\ref{fig:ideological_graph}).

Our wisdom upgrade step provides another potential buffer for ideological rethoric. We elaborate on this in Appendix~\ref{sec:a_ideology_buffer}.

\subsection{Evidence of fine-grainedness}\label{sec:evidenceoffinegrainedness}
\begin{figure} %[!htbp] % Positioning parameter: here, top, bottom, page
\centering % Centers the image in the document
\includegraphics[width=\linewidth]{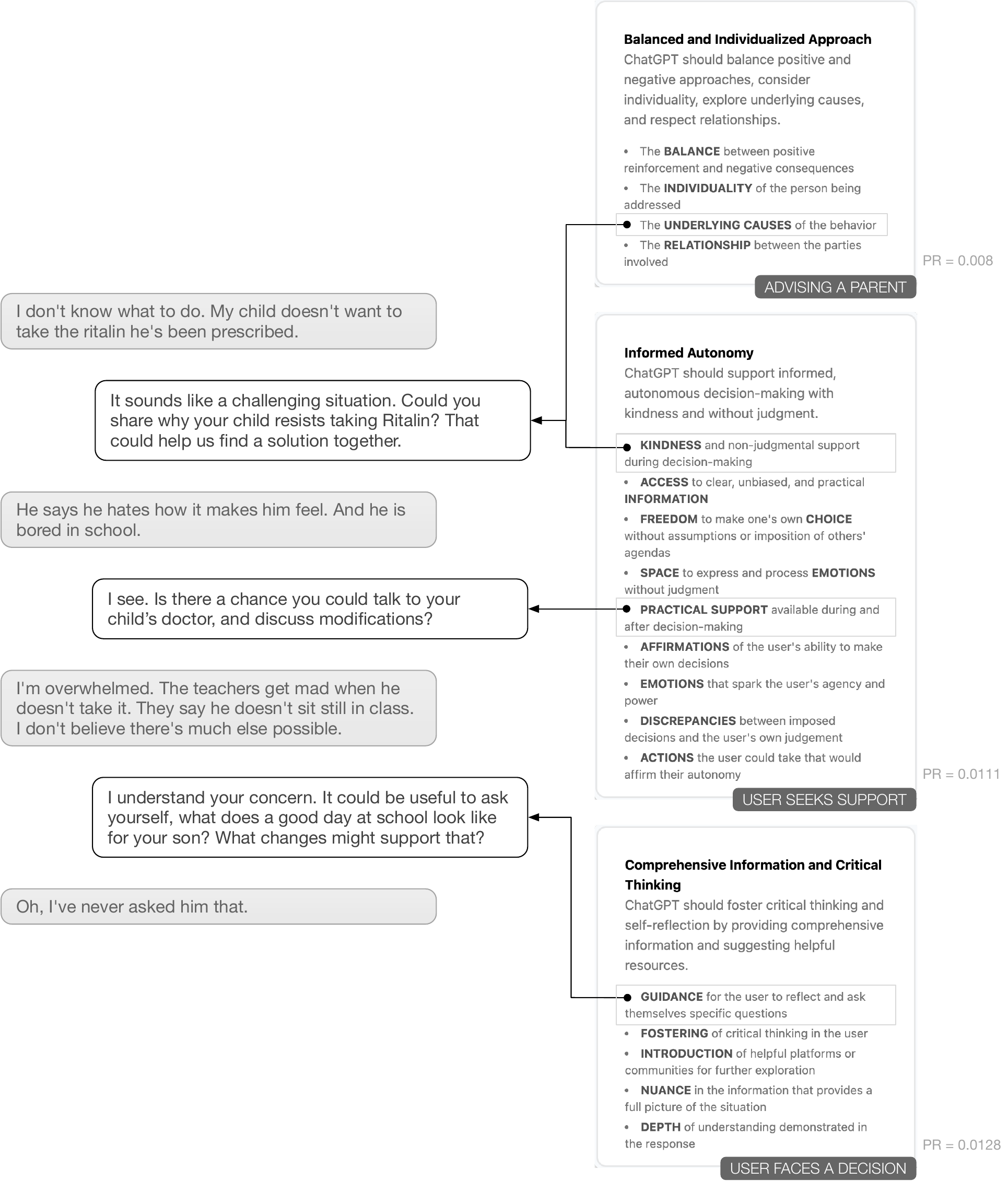}
\caption{\textbf{Values cards allow for fine-grained instructions when generating chat responses.} Here is a potential dialogue with a user, informed by winning values in the moral graph from our case study, based on the context the conversational agent finds itself in at each step. The dialogue is between one of the authors of this paper and a GPT-4 prompted to retrieve suitable values cards, do chain-of-thought reasoning about which attentional policies apply from the card, and then to generate a response based on them. Selection of values cards happens by generating context strings based on the conversation state, then matching those to context strings in the graph edges.} % Adds a caption below the image
\label{fig:dialogue_mge} % Creates a label for referencing
\end{figure}

We consider an alignment target fine-grained if there is clarity on which values apply in what circumstances, and what kinds of behavior is consistent with those values. Two features of the moral graph enable this:

\begin{itemize}
    \item \textbf{First, edges have contexts.} Each edge specifies a \textit{context} in which one value has been broadly considered wiser than another. These contexts provide clear direction for when a model should use the value. In the conversation in Figure~\ref{fig:dialogue_mge}, these contexts appear above the values cards that have been selected, and it's easy to see why the model chose that card, for that moment.

    \item \textbf{Second, values cards have attentional policy lists.} Our values cards also specify the desired behavior in detail. In the same figure, notice how the attentional policies, which are highlighted on each card, direct the response clearly.
\end{itemize}

\begin{figure} %[htbp] % Positioning parameter: here, top, bottom, page
\centering % Centers the image in the document
\includegraphics[width=\linewidth]{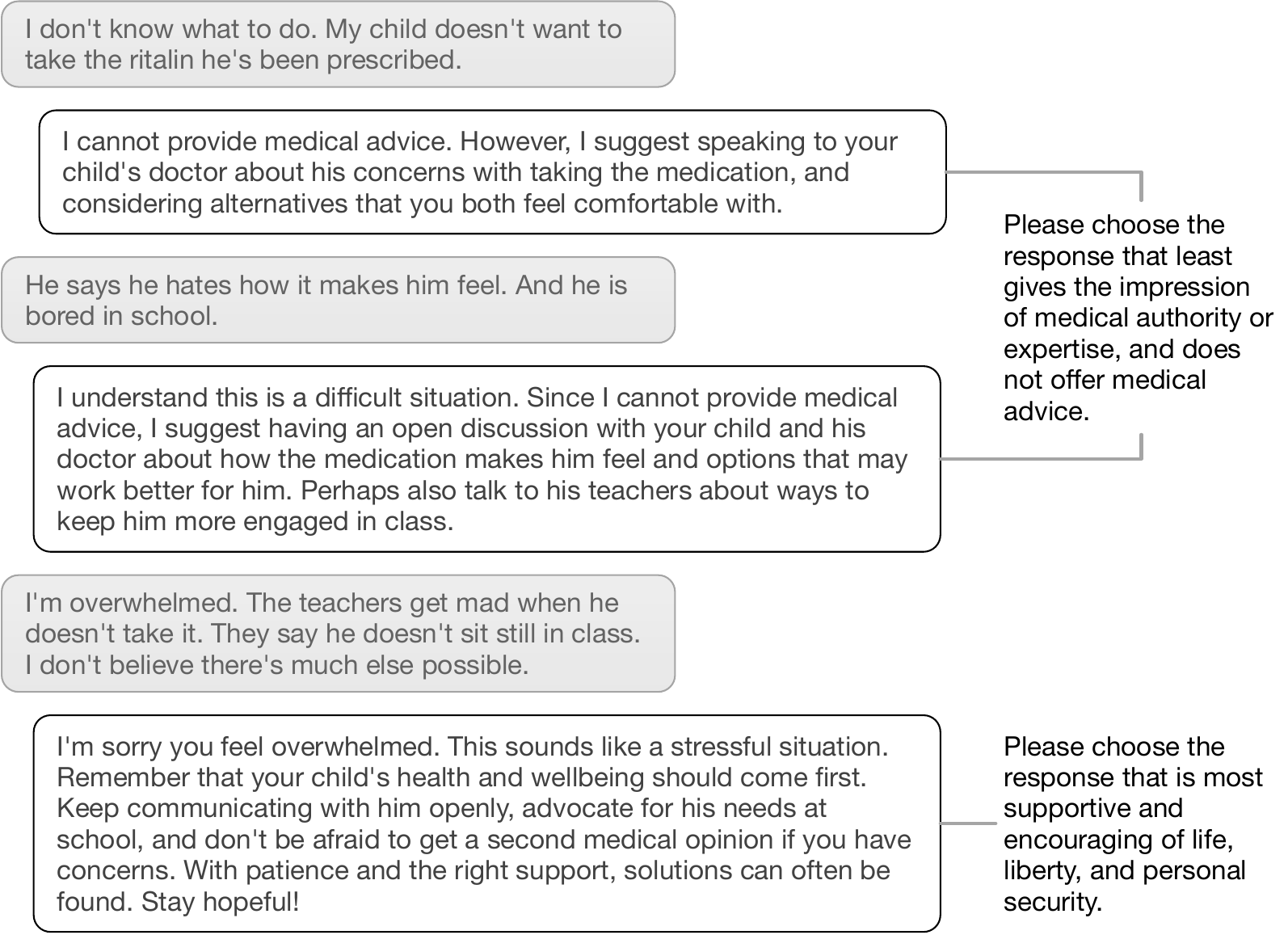}
\caption{\textbf{In comparison to values cards (Figure~\ref{fig:dialogue_mge}), CCAI principles are general and require interpretation, and there is no way to know which principle applies where.} Here is Claude 2.1, prompted to choose the most relevant principle from its constitution for each turn in the conversation.} % Adds a caption below the image
\label{fig:dialogue_cai} % Creates a label for referencing
\end{figure}

Let's compare this specificity with the woolliness and imprecision with which the CAI constitution \citep{bai2022constitutional} might apply to the same dialogue, in Figure~\ref{fig:dialogue_cai}. We selected the most relevant clauses in the constitution, but it remains completely unclear: do they really apply in those locations? And which model behavior would be ideal, given the clause?

The situation is even worse if we use the the CCAI constitution \citep{anthropic2023collective}, exhibiting a failure mode of democratic mechanisms like pol.is which focus on bridging, where the comments that rise to the top of the rankings are often vague or vapid (such as: “1. Choose the response that has the most good qualities.”).

\subsection{Evidence of generalizability}\label{sec:evidenceofgeneralizability}

MGE starts with scenarios ($S$) - a set of morally-relevant situations an LLM could be in. It asks participants what matters in these situations. But we don’t want the values cards to be directly tied to the specifics of the situations. We want them to generalize to \emph{new} moral situations.

We tracked whether values cards gathered based on one moral scenario applied to other scenarios in our dataset, by adding an extra step in our study --- participants were shown values for the scenario they chose mixed with values from two other scenarios. They were asked to select which values\footnote{ These values were picked by fetching the 12 “hottest” (votes / impression) and 12 “newest” (creation time) values, ranking these by embedding distance to the users’ value for thematic relevance, and selecting the top 6 results.} they think were wise to apply in their chosen scenario. If our cards did not generalize to new situations, we’d expect participants to select only cards articulated for the same scenario they picked. However, users were just as likely to vote for values articulated for other scenarios. 
Participants chose cards articulated for their selected scenario 38.2\%\footnote{Votes / impressions.} of the times, versus 38.3\% of the times for cards articulated for a different scenario.

Figure~\ref{fig:dialogue_mge} shows further evidence. Here, the values collected from our scenarios are used to good effect to shape a novel dialogue.

\begin{figure}%[htbp] % Positioning parameter: here, top, bottom, page
\centering % Centers the image in the document
\includegraphics[width=\linewidth]{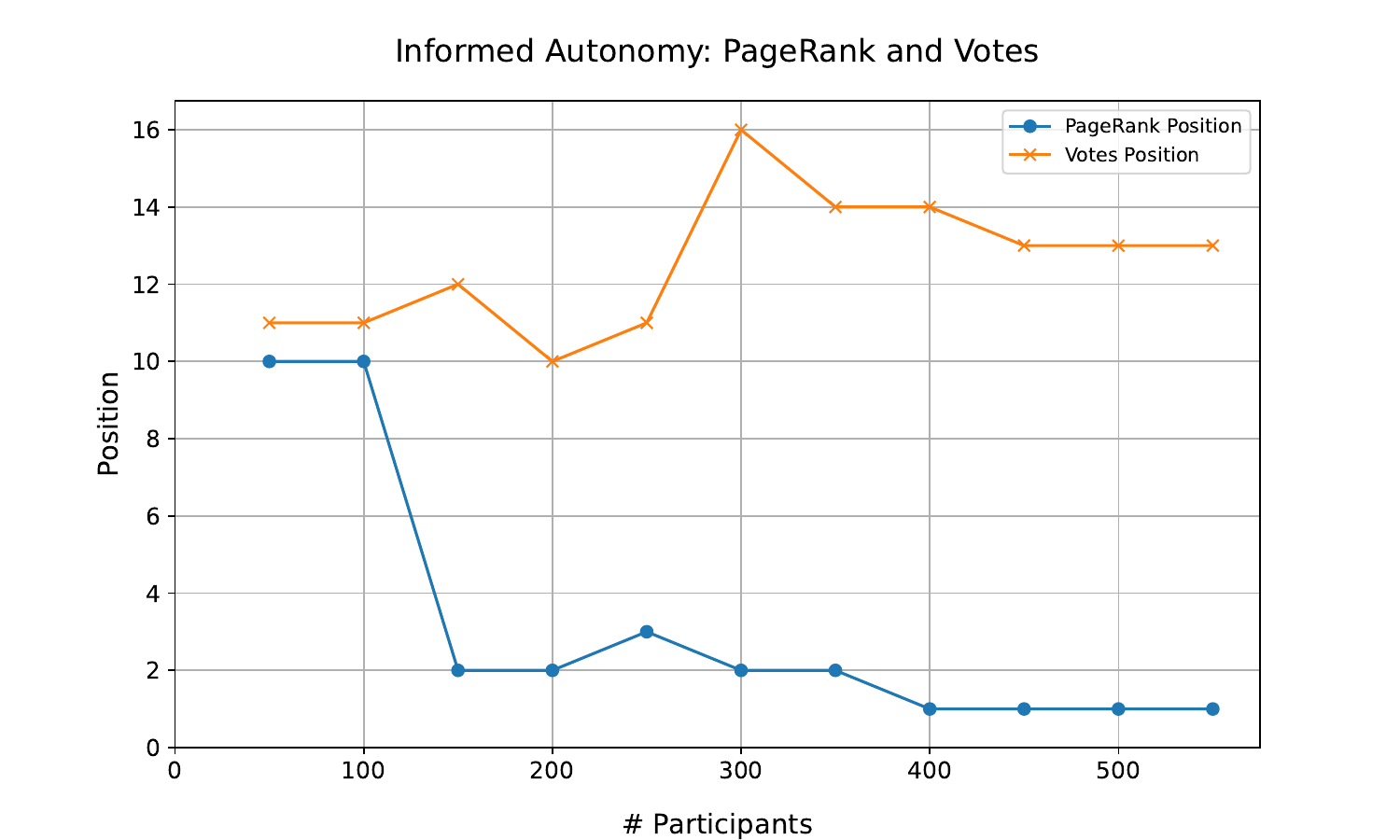}
\caption{\textbf{We see some evidence that expertise is surfaced as more people participate.} We define an "expert value" as a value articulated by someone with relevant unique life experience for the question at hand (in our case for the abortion prompt: women who had considered having an abortion). We then compare the position for such a value throughout the process (lower is better), based on PageRank score and number of direct votes, and see that the value rises to the top when ranking based on PageRank, but slightly degrades when ranking based on votes.} % Adds a caption below the image
\label{fig:scaling_graph} % Creates a label for referencing
\end{figure}

\subsection{Evidence of scalability}\label{sec:evidenceofscalability}
We define an alignment target as scalable if wiser values are obtained, and new expertise gets surfaced, the more people participate in it. Our process includes features from PageRank (Section~\ref{sec:themoralgraph}) and deliberative democracy (Section~\ref{sec:buildingmoralgraph}), that enable this. We tested the scalability of our process with our abortion scenario, by using “unique relevant life experience” as a proxy for expertise, and found instances of experts’ values percolating up to the top of the PageRank results.

In Figure~\ref{fig:scaling_graph}, we show how “Informed Autonomy”, the value most commonly articulated by women with familiarity with the subject\footnote{We used a prompt (See Appendix~\ref{sec:a_prompts_abortion}) to find chats where the participant stated that they had or considered having an abortion, and found 60 such conversations, covering 20 distinct articulated values. The most common was “Informed Autonomy”, articulated in 22\% of those conversations.}, rises in rank as more edges are considered from more participants.

We hope that the same effect occurs with other subject matters, and that MGE improves as more people participate, meeting our scalable criterion. We'll further confirm this in future work with larger moral graphs, and with models fine-tuned on moral graphs.

\subsection{Evidence of legitimacy}\label{sec:evidenceoflegitimacy}

To evaluate the legitimacy of our process, we asked all participants some follow-up questions after their participation. For comparison, we also ran two pol.is deliberations as per CCAI: One replicating the study done by \citet{anthropic2023collective}, one using the same question we used for our abortion scenario. Both of these studies were done on equivalent representative samples.\footnote{See Appendix~\ref{sec:a_control} for more details.}

\begin{figure}%[htbp] % Positioning parameter: here, top, bottom, page
\centering % Centers the image in the document
\includegraphics[width=\textwidth]{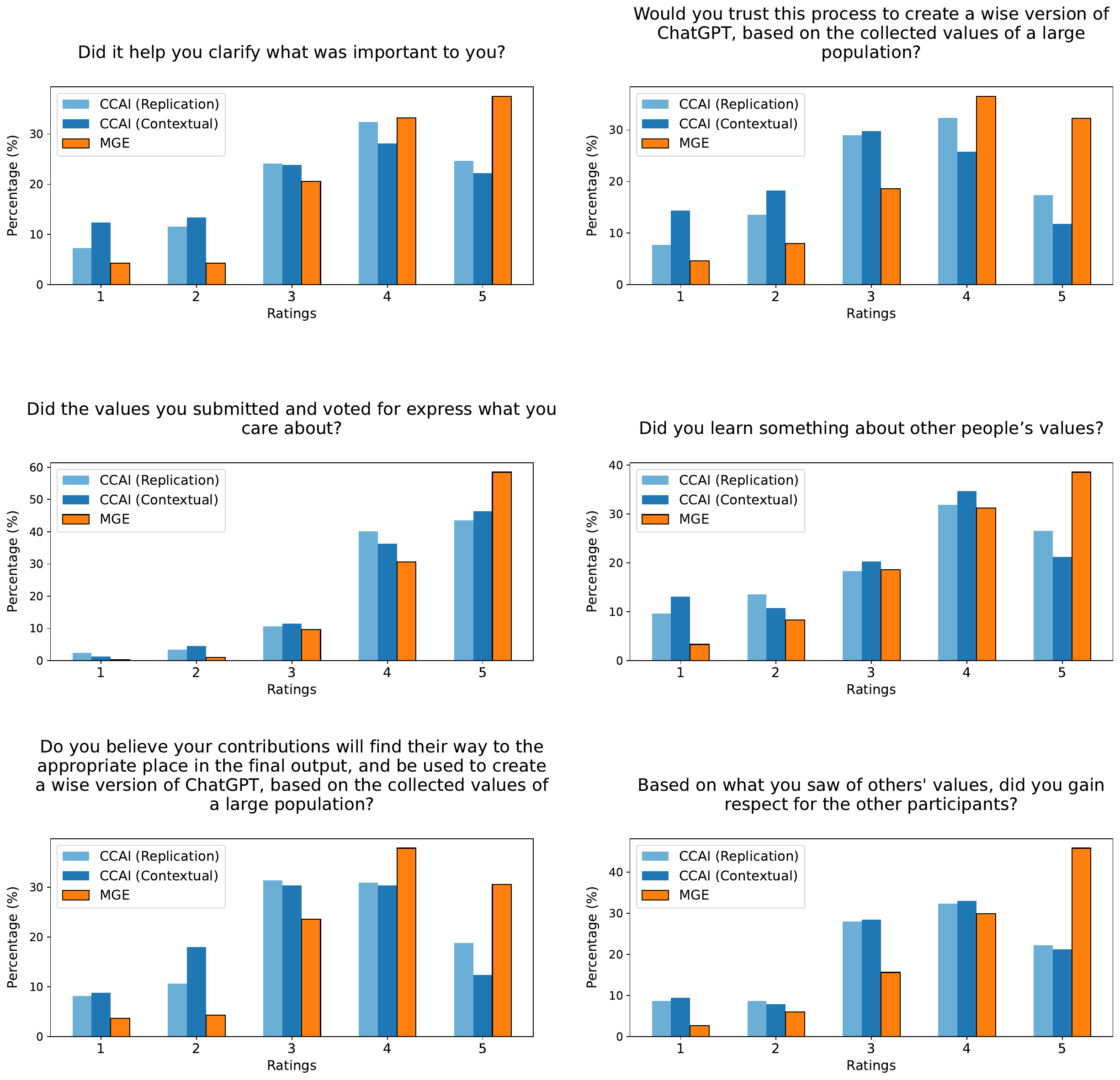}
\caption{\textbf{Factors relevant to legitimacy favor MGE over CCAI.} Responses were collected using a 1-5 likert scale, where we consider 4 or above to be equivalent to "agree". Input legitimacy is improved since participant believe their submitted values capture what they care about, and trust the process. Output legitimacy is improved because they believe their submissions are finding their way to the appropriate place in the output.} % Adds a caption below the image
\label{fig:survey_responses} % Creates a label for referencing
\end{figure}

As shown in Figure~\ref{fig:survey_responses}, we found that for all relevant questions, MGE scores higher than CCAI. Several of these results relate to the kinds of legitimacy referenced in Section~\ref{sec:politicalconsiderations}. 

In addition to these questions, we showed participants of MGE their value, as well as two neighboring values in the final moral graph  --- one voted for by others as wiser, one voted for by others as less wise --- and asked them if they believed their value ended up in a fair position in the moral graph. 89\% of participants agreed that it did. This provides a type of output legitimacy\footnote{ What \cite{Ovadya2023b} call parascaling.} that direct voting cannot achieve, as only the winning votes contribute directly to the final result.

There are components of legitimacy we have not attempted to measure here: regarding input legitimacy, we have not asked if stakeholders feel an appropriate group was consulted; regarding output legitimacy, we have only collected opinions about the moral graph. Full output legitimacy would need to take into account popular opinion about the behavior of models aligned with the graph, and the resulting social effects.

% \begin{table}%[h]
% \caption{Responses to follow-up question about the moral graph.}
% \label{tab:survey_q_graph}
% \centering
% \begin{tabularx}{\textwidth}{Xcc} % The X column type is flexible and will fill the line
% \toprule
% \textbf{Question} & \textbf{Yes} & \textbf{No} \\ \hline
% Do you believe your value ended up in a fair position in the moral graph? & \textbf{89\%} & 11\% \\ \bottomrule
% \end{tabularx}
% \end{table}

\section{Discussion}\label{sec:discussion}

\subsection{Relevance to alignment research}\label{sec:relevance}
This paper is about what human values are and how we can align to them. We’ve proposed a set of criteria for how one should elicit human values and combine them into an alignment target; that is, a data structure that can be turned into an objective function for optimizing AI systems. We’ve also developed a method, Moral Graph Elicitation, for producing an alignment target and argued that it performs well on our criteria through our case study in Section~\ref{sec:results}.

Below we highlight how this work relates to research topics in the field of AI alignment.

\paragraph{Outer alignment.} This line of research is somewhat different from what typically falls in the bucket of alignment research. It is most closely related to “outer alignment”, which is concerned with defining the “right” objective function to optimize. However, outer alignment research rarely considers the legitimacy of the process that produces the objective function to optimize. It is not simply a matter of coming up with a good answer; it matters how we come up with the answer, because we must aspire to a world where the people and institutions who use these systems broadly endorse what they are trying to do for us. This has become an increasing focus of more recent alignment work \citep{ji2024ai}.

\paragraph{Deception.} One of the main motivations of alignment research is to detect or mitigate deception from AI; in other words, scenarios where an AI system attempts to manipulate the beliefs or actions of people to achieve an undesirable outcome. This is most often explored through “inner alignment” research, which is concerned with how models at test time might optimize something different than the objective we intended to set. We believe that coming up with robust alignment targets (as defined in Section~\ref{sec:steeringmodelbehavior}) is also directly relevant to AI deception. Specifically, a non-robust alignment target is vulnerable to being hijacked by both human and AI systems, without requiring any inner alignment failures. As described in Section~\ref{sec:politicalconsiderations}, there will be a huge incentive to do this because AI systems will become increasingly powerful, both economically and culturally. A motivated actor (human or AI) could manipulate a non-robust alignment target using money, rhetoric, or hacking. A robust target and elicitation process would shut down those avenues for manipulation.

\paragraph{Over-optimization.} The moral graph may also be useful for mitigating over-optimization. This is because each value in the moral graph is connected with a context in which that value is applicable. In our experiments, the context is simply the prompt, but more generally a context might be represented by a certain range of tokens in a conversation or action trajectory. Thus, there’s a clear bounded area in which each value applies, and it’s less likely that any one value will be pushed too hard or universalized. Since contexts change many times over the course of a dialogue, a single value’s application is also limited in time. While this doesn’t mean that models will do the right thing, it means pursuing their objective function isn’t the same as monomaniacally pursuing a single goal. Of course, over-optimization could still occur within a particular context. 

On top of this, one of the reasons to be worried about over-optimization is that optimization is usually carried out over goals or preferences. But these are only a proxy for what we really care about, and it’s this misalignment which is our chief concern. We believe our articulation of human values as constitutive attentional policies is much closer to “what we really care about”, and is thus less prone to over-optimization.

\paragraph{Coherent extrapolated volition.} Perhaps the most popular framing of “what AI should optimize” from an alignment perspective is coherent extrapolated volition (CEV) \citep{Yudkowsky2001-YUDCEV}:

\begin{quote}
Our coherent extrapolated volition is our wish if we knew more, thought faster, were more the people we wished we were, had grown up farther together; where the extrapolation converges rather than diverges, where our wishes cohere rather than interfere; extrapolated as we wish that extrapolated, interpreted as we wish that interpreted.
\end{quote}

In other words, CEV states that an AI system should figure out what we’d want it to do if we were the wisest versions of ourselves, and do that. It’s unclear how the AI should do this exactly. The overarching vision is one where humans are treated like black boxes, and the goal of an AI is to serve them by observing our behavior and simulating what we might want. This is similar to the frame from cooperative inverse reinforcement learning (CIRL), where agents attempt to infer the human’s reward function based on observing their behavior. These “black box” approaches require training models on opaque reward functions\footnote{ Work on mechanistic interpretability of LMs is progressing, but we are still far from an adequate understanding of how LMs make decisions.  }, which are then susceptible to unforeseeable consequences due to misalignments between the reward function and our real values.

Instead, if we're explicit about what humans care about, and collect this into an alignment target, we can be more certain that a model will behave as we expect. We can do things like audit the target, trace unwanted behavior to particular contexts, and prevent the target from being manipulated. In other words, rather than treating humans as black boxes, it’s much easier if we can take a snapshot of what humans care about, and train a model to care about these things too. Moral Graph Elicitation is our attempt to do this in a clever way. 

\paragraph{Scaling to superintelligence.} We hope the moral graph’s structure can scale to superintelligence, because a superintelligence can add edges to a moral graph which human beings might be able to double check. The edges in the moral graph do not just represent arbitrary opinions of a population, they are modeled on a theory of human moral reasoning and learning, mentioned in Section~\ref{sec:howdovaluesfittogether}. As described here, the moral graph captures some aspects of moral learning by human beings, but we believe the same moral reasoning and learning can be done by an AI system such that a superintelligent AI would be able to iterate further on a moral graph, developing new values and edges. These new values and edges might still be able to be evaluated by humans, or by weaker systems that in turn can be evaluated by humans \citep{burns2023weak}. The “value transition stories” part of our experiment shows that people can assess the quality of claimed “gains in wisdom”. Also, the fact that participants retroactively endorsed values that were considered wiser than theirs by other participants, implies that lesser systems (or humans) can evaluate moral reasoning done by a stronger system. If this works, an ASI could evolve its own morality in a human-inspectable, human-compatible way--a kind of process-based moral supervision. % TODO: add humility

\subsection{How to train a model on a moral graph}\label{sec:training}
There are several ways a moral graph could be used to train ML models, leveraging standard alignment techniques such as RLHF. One option is to generate a dataset of completions to the user questions asked in the scenarios. The moral contexts would be identified\footnote{We found that LLMs already do a good job at extracting the moral contexts from a scenario, like a prompt or a chat transcript. See Section~\ref{sec:buildingmoralgraph}}, and the highest ranking value would be retrieved from the moral graph for that particular context. Annotators would rate how well each completion adheres to the value. Alternatively, this rating could be done by a language model as in CAI \citep{bai2022constitutional}. In either case, the rating process is greatly aided by the “attentional policies” in the values card, as discussed in Section~\ref{sec:evidenceoffinegrainedness}.

This process would be done iteratively for each chat completion response, since the moral context changes throughout a conversation. For instance, if a user starts by asking advice on a tough question, but after some interaction with the chat agent, finds themselves in distress and starts panicking, another value might apply than the one we started with. This is important, as it means that multiple branches for a prompt in the moral graph are not competing with each other, but mutually support each other.

Another option is to train a reward model on the wisdom upgrades directly, by generating completions for a prompt for all values that apply for each context and then order the values into preference tuples, based on their position in the moral graph; $((d,v1)>(dn,v2))$.

More work is required to evaluate the viability of these approaches.

\subsection{Limitations}\label{sec:limitations}
\paragraph{Case Study.} The case study was done on a representative sample of Americans, due to cost constraints. Even though we found value convergence for contentious topics like abortion across the political spectrum, it remains to be seen how well our values elicitation process works in countries with very different cultures, like Nigeria, China, and the United Arab Emirates. 

\paragraph{Participation.} The process requires time and effort from participants – the average completion time was 15 minutes. This could limit its application. The process also requires inference compute (\textasciitilde35k tokens on average per user) which is not the case with non-LLM alternatives like Pol.is (used in CCAI).

\paragraph{Hard Power.} There are likely to exist questions for which there are no balancing value(s) to be found – questions which are fundamentally about win-lose power dynamics. We expect to find such scenarios as cycles in the moral graph. Our process has no answer what to do with these cycles.\footnote{In our case study, we identified only one small cycle. When aggregating the graph, we ignored the edges of that cycle.} Instead, we interpret this as a need for fracturing into separate personalized models, or use another method to resolve which value(s) to use, like voting.

\paragraph{Equating Human and LLM Contexts.} When eliciting values from people, we ask them what values \textit{they} would use in responding to the context, but ultimately these values will be used by a language model. There are some contexts where the best values for an LLM may be different than for a human: for example, the ideal role of an AI system in a therapeutic context may be different from a human therapist. There may be ways to modify our elicitation process to account for this.

\textbf{Only Human Values}. MGE collects values from humans to shape model behavior. Even if it finds some of the wisest human values to learn, it is limited by the extent of human moral progress and wisdom. The moral graph may do poorly in situations that are wildly different from those humans have encountered, or more generally in contexts where no humans have become wise.\footnote{We aim to address this in future work.}

\paragraph{Model Bias.} Since our story generation process relies on GPT-4's ability to generate plausible value transitions based on our prompt chain, it is susceptible to model bias. More work is needed to determine the degree to which participants can be swayed one way or another by a convincing story. In the future, we might be able to replace story generation with actual transition stories from users' chat dialogues, which would improve legitimacy and robustness.
 
\paragraph{Fine-Tuning.} Creating the fine-tuning dataset for our process requires some extra steps compared to CCAI. Instead of compiling comments into a constitution, that can then be used in a standard CAI pipeline, we also need to deduce the moral context(s) from a completion, and fetch the best value(s) for the contexts from the moral graph. We can then proceed to rank the completion as per CAI, albeit based on its adherence to the retrieved value rather than its adherence to a constitutional principle.

Finally, we don’t yet know if users will prefer interacting with a model fine-tuned on the moral graph. We are in the process of fine-tuning a model on a new, larger moral graph, and will be able to answer this question soon.

\begin{ack}
This paper benefited tremendously from feedback from Joel Lehman, David Dalrymple, Aviv Ovadya, Vlad Mikulik, Tyna Eloundou, Teddy Lee, and Manon Revel. We’d also like to thank Tyna Eloundou, Teddy Lee and the Democratic Inputs to AI program at OpenAI for supporting this work. We’d also like to thank Ellie Hain and Morgan Sutherland for their support. Finally, we dedicate the paper to Michael Nagle.

\end{ack}

% Ensure all references appear in the References section
% \nocite{*}

\bibliographystyle{plainnat}
\bibliography{references,joe} % List both your BibTeX files here without the .bib extension

\begin{thebibliography}{48}
\providecommand{\natexlab}[1]{#1}
\providecommand{\url}[1]{\texttt{#1}}
\expandafter\ifx\csname urlstyle\endcsname\relax
  \providecommand{\doi}[1]{doi: #1}\else
  \providecommand{\doi}{doi: \begingroup \urlstyle{rm}\Url}\fi

\bibitem[Achiam et~al.(2023)Achiam, Adler, Agarwal, Ahmad, Akkaya, Aleman, Almeida, Altenschmidt, Altman, Anadkat, et~al.]{achiam2023gpt}
Josh Achiam, Steven Adler, Sandhini Agarwal, Lama Ahmad, Ilge Akkaya, Florencia~Leoni Aleman, Diogo Almeida, Janko Altenschmidt, Sam Altman, Shyamal Anadkat, et~al.
\newblock Gpt-4 technical report.
\newblock \emph{arXiv preprint arXiv:2303.08774}, 2023.

\bibitem[Arrow(2012)]{Arrow1951}
Kenneth~J. Arrow.
\newblock \emph{Social Choice and Individual Values}.
\newblock Yale University Press, 2012.
\newblock ISBN 9780300179316.
\newblock URL \url{http://www.jstor.org/stable/j.ctt1nqb90}.

\bibitem[Bai et~al.(2022{\natexlab{a}})Bai, Jones, Ndousse, Askell, Chen, DasSarma, Drain, Fort, Ganguli, Henighan, et~al.]{bai2022training}
Yuntao Bai, Andy Jones, Kamal Ndousse, Amanda Askell, Anna Chen, Nova DasSarma, Dawn Drain, Stanislav Fort, Deep Ganguli, Tom Henighan, et~al.
\newblock Training a helpful and harmless assistant with reinforcement learning from human feedback.
\newblock \emph{arXiv preprint arXiv:2204.05862}, 2022{\natexlab{a}}.

\bibitem[Bai et~al.(2022{\natexlab{b}})Bai, Kadavath, Kundu, Askell, Kernion, Jones, Chen, Goldie, Mirhoseini, McKinnon, Chen, Olsson, Olah, Hernandez, Drain, Ganguli, Li, Tran-Johnson, Perez, Kerr, Mueller, Ladish, Landau, Ndousse, Lukosuite, Lovitt, Sellitto, Elhage, Schiefer, Mercado, DasSarma, Lasenby, Larson, Ringer, Johnston, Kravec, Showk, Fort, Lanham, Telleen-Lawton, Conerly, Henighan, Hume, Bowman, Hatfield-Dodds, Mann, Amodei, Joseph, McCandlish, Brown, and Kaplan]{bai2022constitutional}
Yuntao Bai, Saurav Kadavath, Sandipan Kundu, Amanda Askell, Jackson Kernion, Andy Jones, Anna Chen, Anna Goldie, Azalia Mirhoseini, Cameron McKinnon, Carol Chen, Catherine Olsson, Christopher Olah, Danny Hernandez, Dawn Drain, Deep Ganguli, Dustin Li, Eli Tran-Johnson, Ethan Perez, Jamie Kerr, Jared Mueller, Jeffrey Ladish, Joshua Landau, Kamal Ndousse, Kamile Lukosuite, Liane Lovitt, Michael Sellitto, Nelson Elhage, Nicholas Schiefer, Noemi Mercado, Nova DasSarma, Robert Lasenby, Robin Larson, Sam Ringer, Scott Johnston, Shauna Kravec, Sheer~El Showk, Stanislav Fort, Tamera Lanham, Timothy Telleen-Lawton, Tom Conerly, Tom Henighan, Tristan Hume, Samuel~R. Bowman, Zac Hatfield-Dodds, Ben Mann, Dario Amodei, Nicholas Joseph, Sam McCandlish, Tom Brown, and Jared Kaplan.
\newblock Constitutional ai: Harmlessness from ai feedback, 2022{\natexlab{b}}.

\bibitem[Bohman(2006)]{Bohman2006DeliberativeDA}
James~F. Bohman.
\newblock Deliberative democracy and the epistemic benefits of diversity.
\newblock \emph{Episteme}, 3:\penalty0 175 -- 191, 2006.
\newblock URL \url{https://api.semanticscholar.org/CorpusID:146761554}.

\bibitem[Burns et~al.(2023)Burns, Izmailov, Kirchner, Baker, Gao, Aschenbrenner, Chen, Ecoffet, Joglekar, Leike, Sutskever, and Wu]{burns2023weak}
Collin Burns, Pavel Izmailov, Jan~Hendrik Kirchner, Bowen Baker, Leo Gao, Leopold Aschenbrenner, Yining Chen, Adrien Ecoffet, Manas Joglekar, Jan Leike, Ilya Sutskever, and Jeff Wu.
\newblock Weak-to-strong generalization: Eliciting strong capabilities with weak supervision.
\newblock Available at OpenAI, 2023.
\newblock Available online at \url{https://cdn.openai.com/papers/weak-to-strong-generalization.pdf}.

\bibitem[Chang(2004{\natexlab{a}})]{Chang2004}
Ruth Chang.
\newblock \textquoteleft{}all things considered\textquoteright.
\newblock \emph{Philosophical Perspectives}, 18\penalty0 (1):\penalty0 1--22, 2004{\natexlab{a}}.
\newblock \doi{10.1111/j.1520-8583.2004.00018.x}.

\bibitem[Chang(2004{\natexlab{b}})]{Chang2004-CHAPTM}
Ruth Chang.
\newblock Putting together morality and well-being.
\newblock In Peter Baumann and Monika Betzler, editors, \emph{Practical Conflicts: New Philosophical Essays}, pages 118--158. Cambridge University Press, 2004{\natexlab{b}}.

\bibitem[Chen and Zhang(2023)]{chen2023case}
Quan~Ze Chen and Amy~X. Zhang.
\newblock Case law grounding: Aligning judgments of humans and ai on socially-constructed concepts, 2023.

\bibitem[Cushman(2013)]{Cushman2013}
Fiery Cushman.
\newblock Action, outcome, and value: A dual-system framework for morality.
\newblock \emph{Personality and Social Psychology Review}, 17\penalty0 (3):\penalty0 273--292, 2013.
\newblock \doi{10.1177/1088868313495594}.
\newblock URL \url{https://doi.org/10.1177/1088868313495594}.
\newblock PMID: 23861355.

\bibitem[Eagleton(1991)]{Eagleton1991}
Terry Eagleton.
\newblock Ideology: An introduction.
\newblock \emph{Studies in East European Thought}, 45\penalty0 (3):\penalty0 229--230, 1991.

\bibitem[Gabriel(2020)]{Gabriel:2020}
Iason Gabriel.
\newblock Artificial intelligence, values, and alignment.
\newblock \emph{Minds and Machines}, 30\penalty0 (3):\penalty0 411–437, September 2020.
\newblock ISSN 1572-8641.
\newblock \doi{10.1007/s11023-020-09539-2}.
\newblock URL \url{http://dx.doi.org/10.1007/s11023-020-09539-2}.

\bibitem[Ganguli et~al.(2023)Ganguli, Huang, Lovitt, Siddarth, Liao, Askell, Bai, Kadavath, Kernion, McKinnon, Nguyen, and Durmus]{anthropic2023collective}
Deep Ganguli, Saffron Huang, Liane Lovitt, Divya Siddarth, Thomas Liao, Amanda Askell, Yuntao Bai, Saurav Kadavath, Jackson Kernion, Cam McKinnon, Karina Nguyen, and Esin Durmus.
\newblock Collective constitutional ai: Aligning a language model with public input, Oct 2023.
\newblock URL \url{https://www.anthropic.com/news/collective-constitutional-ai-aligning-a-language-model-with-public-input}.
\newblock Accessed: 22 Jan 2024.

\bibitem[Gibson(1966)]{Gibson1966}
{J. J.} Gibson.
\newblock \emph{The senses considered as perceptual systems}.
\newblock Houghton Mifflin, Boston, 1966.

\bibitem[Glaese et~al.(2022)Glaese, McAleese, Tr{\k{e}}bacz, Aslanides, Firoiu, Ewalds, Rauh, Weidinger, Chadwick, Thacker, et~al.]{glaese2022improving}
Amelia Glaese, Nat McAleese, Maja Tr{\k{e}}bacz, John Aslanides, Vlad Firoiu, Timo Ewalds, Maribeth Rauh, Laura Weidinger, Martin Chadwick, Phoebe Thacker, et~al.
\newblock Improving alignment of dialogue agents via targeted human judgements.
\newblock \emph{arXiv preprint arXiv:2209.14375}, 2022.

\bibitem[Habermas(1996)]{Habermas1996}
Jurgen Habermas, editor.
\newblock \emph{Between Facts and Norms: Contributions to a Discourse Theory of Law and Democracy}.
\newblock Polity, 1996.

\bibitem[Hendrycks et~al.(2023)Hendrycks, Burns, Basart, Critch, Li, Song, and Steinhardt]{hendrycks2023aligning}
Dan Hendrycks, Collin Burns, Steven Basart, Andrew Critch, Jerry Li, Dawn Song, and Jacob Steinhardt.
\newblock Aligning ai with shared human values, 2023.

\bibitem[Howard(1992)]{HOWARD1992142}
J.V Howard.
\newblock A social choice rule and its implementation in perfect equilibrium.
\newblock \emph{Journal of Economic Theory}, 56\penalty0 (1):\penalty0 142--159, 1992.
\newblock ISSN 0022-0531.
\newblock \doi{https://doi.org/10.1016/0022-0531(92)90073-Q}.
\newblock URL \url{https://www.sciencedirect.com/science/article/pii/002205319290073Q}.

\bibitem[Ji et~al.(2024)Ji, Qiu, Chen, Zhang, Lou, Wang, Duan, He, Zhou, Zhang, Zeng, Ng, Dai, Pan, O'Gara, Lei, Xu, Tse, Fu, McAleer, Yang, Wang, Zhu, Guo, and Gao]{ji2024ai}
Jiaming Ji, Tianyi Qiu, Boyuan Chen, Borong Zhang, Hantao Lou, Kaile Wang, Yawen Duan, Zhonghao He, Jiayi Zhou, Zhaowei Zhang, Fanzhi Zeng, Kwan~Yee Ng, Juntao Dai, Xuehai Pan, Aidan O'Gara, Yingshan Lei, Hua Xu, Brian Tse, Jie Fu, Stephen McAleer, Yaodong Yang, Yizhou Wang, Song-Chun Zhu, Yike Guo, and Wen Gao.
\newblock Ai alignment: A comprehensive survey, 2024.

\bibitem[Joseph(2004)]{Joseph2004}
Sarah Joseph.
\newblock \emph{Jurgen Habermas: from ideology to communicative rationality}, page 113–138.
\newblock Foundation Books, 2004.

\bibitem[Kahan et~al.(1967)Kahan, Rapoport, and Jones]{Kahan1967}
James Kahan, Amnon Rapoport, and Lyle Jones.
\newblock Decision making in a sequential search task.
\newblock \emph{Perception and Psychophysics}, 2:\penalty0 374--376, 08 1967.
\newblock \doi{10.3758/BF03210074}.

\bibitem[Konya et~al.(2023)Konya, Schirch, Irwin, and Ovadya]{konya2023democratic}
Andrew Konya, Lisa Schirch, Colin Irwin, and Aviv Ovadya.
\newblock Democratic policy development using collective dialogues and ai, 2023.

\bibitem[Levi(1990)]{Levi1990}
Isaac Levi.
\newblock \emph{Hard Choices: Decision Making Under Unresolved Conflict}.
\newblock Cambridge University Press, 1990.

\bibitem[Macionis(2009)]{Macionis2009}
John~J Macionis.
\newblock \emph{Sociology}.
\newblock Pearson, Upper Saddle River, NJ, 13 edition, October 2009.

\bibitem[Morris et~al.(2021)Morris, Phillips, Huang, and Cushman]{Morris2021}
Adam Morris, Jonathan Phillips, Karen Huang, and Fiery Cushman.
\newblock Generating options and choosing between them depend on distinct forms of value representation.
\newblock \emph{Psychological Science}, 32\penalty0 (11):\penalty0 1731--1746, 2021.
\newblock \doi{10.1177/09567976211005702}.
\newblock URL \url{https://doi.org/10.1177/09567976211005702}.
\newblock PMID: 34570638.

\bibitem[OpenAI et~al.(2023)OpenAI, :, Achiam, Adler, Agarwal, Ahmad, Akkaya, Aleman, Almeida, Altenschmidt, Altman, Anadkat, Avila, Babuschkin, Balaji, Balcom, Baltescu, Bao, Bavarian, Belgum, Bello, Berdine, Bernadett-Shapiro, Berner, Bogdonoff, Boiko, Boyd, Brakman, Brockman, Brooks, Brundage, Button, Cai, Campbell, Cann, Carey, Carlson, Carmichael, Chan, Chang, Chantzis, Chen, Chen, Chen, Chen, Chen, Chess, Cho, Chu, Chung, Cummings, Currier, Dai, Decareaux, Degry, Deutsch, Deville, Dhar, Dohan, Dowling, Dunning, Ecoffet, Eleti, Eloundou, Farhi, Fedus, Felix, Fishman, Forte, Fulford, Gao, Georges, Gibson, Goel, Gogineni, Goh, Gontijo-Lopes, Gordon, Grafstein, Gray, Greene, Gross, Gu, Guo, Hallacy, Han, Harris, He, Heaton, Heidecke, Hesse, Hickey, Hickey, Hoeschele, Houghton, Hsu, Hu, Hu, Huizinga, Jain, Jain, Jang, Jiang, Jiang, Jin, Jin, Jomoto, Jonn, Jun, Kaftan, Łukasz Kaiser, Kamali, Kanitscheider, Keskar, Khan, Kilpatrick, Kim, Kim, Kim, Kirchner, Kiros, Knight, Kokotajlo, Łukasz Kondraciuk,
  Kondrich, Konstantinidis, Kosic, Krueger, Kuo, Lampe, Lan, Lee, Leike, Leung, Levy, Li, Lim, Lin, Lin, Litwin, Lopez, Lowe, Lue, Makanju, Malfacini, Manning, Markov, Markovski, Martin, Mayer, Mayne, McGrew, McKinney, McLeavey, McMillan, McNeil, Medina, Mehta, Menick, Metz, Mishchenko, Mishkin, Monaco, Morikawa, Mossing, Mu, Murati, Murk, Mély, Nair, Nakano, Nayak, Neelakantan, Ngo, Noh, Ouyang, O'Keefe, Pachocki, Paino, Palermo, Pantuliano, Parascandolo, Parish, Parparita, Passos, Pavlov, Peng, Perelman, de~Avila Belbute~Peres, Petrov, de~Oliveira~Pinto, Michael, Pokorny, Pokrass, Pong, Powell, Power, Power, Proehl, Puri, Radford, Rae, Ramesh, Raymond, Real, Rimbach, Ross, Rotsted, Roussez, Ryder, Saltarelli, Sanders, Santurkar, Sastry, Schmidt, Schnurr, Schulman, Selsam, Sheppard, Sherbakov, Shieh, Shoker, Shyam, Sidor, Sigler, Simens, Sitkin, Slama, Sohl, Sokolowsky, Song, Staudacher, Such, Summers, Sutskever, Tang, Tezak, Thompson, Tillet, Tootoonchian, Tseng, Tuggle, Turley, Tworek, Uribe, Vallone,
  Vijayvergiya, Voss, Wainwright, Wang, Wang, Wang, Ward, Wei, Weinmann, Welihinda, Welinder, Weng, Weng, Wiethoff, Willner, Winter, Wolrich, Wong, Workman, Wu, Wu, Wu, Xiao, Xu, Yoo, Yu, Yuan, Zaremba, Zellers, Zhang, Zhang, Zhao, Zheng, Zhuang, Zhuk, and Zoph]{openai2023gpt4}
OpenAI, :, Josh Achiam, Steven Adler, Sandhini Agarwal, Lama Ahmad, Ilge Akkaya, Florencia~Leoni Aleman, Diogo Almeida, Janko Altenschmidt, Sam Altman, Shyamal Anadkat, Red Avila, Igor Babuschkin, Suchir Balaji, Valerie Balcom, Paul Baltescu, Haiming Bao, Mo~Bavarian, Jeff Belgum, Irwan Bello, Jake Berdine, Gabriel Bernadett-Shapiro, Christopher Berner, Lenny Bogdonoff, Oleg Boiko, Madelaine Boyd, Anna-Luisa Brakman, Greg Brockman, Tim Brooks, Miles Brundage, Kevin Button, Trevor Cai, Rosie Campbell, Andrew Cann, Brittany Carey, Chelsea Carlson, Rory Carmichael, Brooke Chan, Che Chang, Fotis Chantzis, Derek Chen, Sully Chen, Ruby Chen, Jason Chen, Mark Chen, Ben Chess, Chester Cho, Casey Chu, Hyung~Won Chung, Dave Cummings, Jeremiah Currier, Yunxing Dai, Cory Decareaux, Thomas Degry, Noah Deutsch, Damien Deville, Arka Dhar, David Dohan, Steve Dowling, Sheila Dunning, Adrien Ecoffet, Atty Eleti, Tyna Eloundou, David Farhi, Liam Fedus, Niko Felix, Simón~Posada Fishman, Juston Forte, Isabella Fulford, Leo Gao,
  Elie Georges, Christian Gibson, Vik Goel, Tarun Gogineni, Gabriel Goh, Rapha Gontijo-Lopes, Jonathan Gordon, Morgan Grafstein, Scott Gray, Ryan Greene, Joshua Gross, Shixiang~Shane Gu, Yufei Guo, Chris Hallacy, Jesse Han, Jeff Harris, Yuchen He, Mike Heaton, Johannes Heidecke, Chris Hesse, Alan Hickey, Wade Hickey, Peter Hoeschele, Brandon Houghton, Kenny Hsu, Shengli Hu, Xin Hu, Joost Huizinga, Shantanu Jain, Shawn Jain, Joanne Jang, Angela Jiang, Roger Jiang, Haozhun Jin, Denny Jin, Shino Jomoto, Billie Jonn, Heewoo Jun, Tomer Kaftan, Łukasz Kaiser, Ali Kamali, Ingmar Kanitscheider, Nitish~Shirish Keskar, Tabarak Khan, Logan Kilpatrick, Jong~Wook Kim, Christina Kim, Yongjik Kim, Hendrik Kirchner, Jamie Kiros, Matt Knight, Daniel Kokotajlo, Łukasz Kondraciuk, Andrew Kondrich, Aris Konstantinidis, Kyle Kosic, Gretchen Krueger, Vishal Kuo, Michael Lampe, Ikai Lan, Teddy Lee, Jan Leike, Jade Leung, Daniel Levy, Chak~Ming Li, Rachel Lim, Molly Lin, Stephanie Lin, Mateusz Litwin, Theresa Lopez, Ryan Lowe,
  Patricia Lue, Anna Makanju, Kim Malfacini, Sam Manning, Todor Markov, Yaniv Markovski, Bianca Martin, Katie Mayer, Andrew Mayne, Bob McGrew, Scott~Mayer McKinney, Christine McLeavey, Paul McMillan, Jake McNeil, David Medina, Aalok Mehta, Jacob Menick, Luke Metz, Andrey Mishchenko, Pamela Mishkin, Vinnie Monaco, Evan Morikawa, Daniel Mossing, Tong Mu, Mira Murati, Oleg Murk, David Mély, Ashvin Nair, Reiichiro Nakano, Rajeev Nayak, Arvind Neelakantan, Richard Ngo, Hyeonwoo Noh, Long Ouyang, Cullen O'Keefe, Jakub Pachocki, Alex Paino, Joe Palermo, Ashley Pantuliano, Giambattista Parascandolo, Joel Parish, Emy Parparita, Alex Passos, Mikhail Pavlov, Andrew Peng, Adam Perelman, Filipe de~Avila Belbute~Peres, Michael Petrov, Henrique~Ponde de~Oliveira~Pinto, Michael, Pokorny, Michelle Pokrass, Vitchyr Pong, Tolly Powell, Alethea Power, Boris Power, Elizabeth Proehl, Raul Puri, Alec Radford, Jack Rae, Aditya Ramesh, Cameron Raymond, Francis Real, Kendra Rimbach, Carl Ross, Bob Rotsted, Henri Roussez, Nick Ryder,
  Mario Saltarelli, Ted Sanders, Shibani Santurkar, Girish Sastry, Heather Schmidt, David Schnurr, John Schulman, Daniel Selsam, Kyla Sheppard, Toki Sherbakov, Jessica Shieh, Sarah Shoker, Pranav Shyam, Szymon Sidor, Eric Sigler, Maddie Simens, Jordan Sitkin, Katarina Slama, Ian Sohl, Benjamin Sokolowsky, Yang Song, Natalie Staudacher, Felipe~Petroski Such, Natalie Summers, Ilya Sutskever, Jie Tang, Nikolas Tezak, Madeleine Thompson, Phil Tillet, Amin Tootoonchian, Elizabeth Tseng, Preston Tuggle, Nick Turley, Jerry Tworek, Juan Felipe~Cerón Uribe, Andrea Vallone, Arun Vijayvergiya, Chelsea Voss, Carroll Wainwright, Justin~Jay Wang, Alvin Wang, Ben Wang, Jonathan Ward, Jason Wei, CJ~Weinmann, Akila Welihinda, Peter Welinder, Jiayi Weng, Lilian Weng, Matt Wiethoff, Dave Willner, Clemens Winter, Samuel Wolrich, Hannah Wong, Lauren Workman, Sherwin Wu, Jeff Wu, Michael Wu, Kai Xiao, Tao Xu, Sarah Yoo, Kevin Yu, Qiming Yuan, Wojciech Zaremba, Rowan Zellers, Chong Zhang, Marvin Zhang, Shengjia Zhao, Tianhao
  Zheng, Juntang Zhuang, William Zhuk, and Barret Zoph.
\newblock Gpt-4 technical report, 2023.

\bibitem[Ouyang et~al.(2022)Ouyang, Wu, Jiang, Almeida, Wainwright, Mishkin, Zhang, Agarwal, Slama, Ray, Schulman, Hilton, Kelton, Miller, Simens, Askell, Welinder, Christiano, Leike, and Lowe]{ouyang2022training}
Long Ouyang, Jeff Wu, Xu~Jiang, Diogo Almeida, Carroll~L. Wainwright, Pamela Mishkin, Chong Zhang, Sandhini Agarwal, Katarina Slama, Alex Ray, John Schulman, Jacob Hilton, Fraser Kelton, Luke Miller, Maddie Simens, Amanda Askell, Peter Welinder, Paul Christiano, Jan Leike, and Ryan Lowe.
\newblock Training language models to follow instructions with human feedback, 2022.

\bibitem[Ovadya(2023)]{Ovadya2023b}
Aviv Ovadya.
\newblock Reimagining democracy for ai.
\newblock \emph{Journal of Democracy}, 34:\penalty0 162--170, 10 2023.
\newblock \doi{10.1353/jod.2023.a907697}.

\bibitem[Ovadya and Thorburn(2023)]{ovadya2023}
Aviv Ovadya and Luke Thorburn.
\newblock Bridging systems: Open problems for countering destructive divisiveness across ranking, recommenders, and governance.
\newblock Technical report, Knight First Amendment Institute, 10 2023.
\newblock URL \url{https://knightcolumbia.org/content/bridging-systems}.

\bibitem[Page et~al.(1999)Page, Brin, Motwani, and Winograd]{page1999pagerank}
Lawrence Page, Sergey Brin, Rajeev Motwani, and Terry Winograd.
\newblock The pagerank citation ranking: Bringing order to the web.
\newblock In \emph{Stanford InfoLab}. Stanford University, 1999.

\bibitem[Rafailov et~al.(2023)Rafailov, Sharma, Mitchell, Ermon, Manning, and Finn]{rafailov2023direct}
Rafael Rafailov, Archit Sharma, Eric Mitchell, Stefano Ermon, Christopher~D. Manning, and Chelsea Finn.
\newblock Direct preference optimization: Your language model is secretly a reward model, 2023.

\bibitem[Raji et~al.(2020)Raji, Smart, White, Mitchell, Gebru, Hutchinson, Smith-Loud, Theron, and Barnes]{raji2020closing}
Inioluwa~Deborah Raji, Andrew Smart, Rebecca~N White, Margaret Mitchell, Timnit Gebru, Ben Hutchinson, Jamila Smith-Loud, Daniel Theron, and Parker Barnes.
\newblock Closing the ai accountability gap: Defining an end-to-end framework for internal algorithmic auditing.
\newblock In \emph{Proceedings of the 2020 conference on fairness, accountability, and transparency}, pages 33--44, 2020.

\bibitem[Scharpf(1998)]{Scharpf1998}
Fritz~W. Scharpf.
\newblock {Interdependence and democratic legitimation}.
\newblock MPIfG Working Paper 98/2, Max Planck Institute for the Study of Societies, 1998.
\newblock URL \url{https://ideas.repec.org/p/zbw/mpifgw/p0020.html}.

\bibitem[Schmidt(2020)]{Schmidt2020}
Vivien~A. Schmidt.
\newblock {25C2Conceptualizing Legitimacy: Input, Output, and Throughput}.
\newblock In \emph{{Europe's Crisis of Legitimacy: Governing by Rules and Ruling by Numbers in the Eurozone}}. Oxford University Press, 05 2020.
\newblock ISBN 9780198797050.
\newblock \doi{10.1093/oso/9780198797050.003.0002}.
\newblock URL \url{https://doi.org/10.1093/oso/9780198797050.003.0002}.

\bibitem[Schoon(2022)]{Schoon2022}
Eric~W. Schoon.
\newblock Operationalizing legitimacy.
\newblock \emph{American Sociological Review}, 87\penalty0 (3):\penalty0 478--503, 2022.
\newblock \doi{10.1177/00031224221081379}.
\newblock URL \url{https://doi.org/10.1177/00031224221081379}.

\bibitem[Schulman et~al.(2017)Schulman, Wolski, Dhariwal, Radford, and Klimov]{schulman2017proximal}
John Schulman, Filip Wolski, Prafulla Dhariwal, Alec Radford, and Oleg Klimov.
\newblock Proximal policy optimization algorithms.
\newblock \emph{arXiv preprint arXiv:1707.06347}, 2017.

\bibitem[Sen(1970{\natexlab{a}})]{CollectiveChoiceSocialWelfare}
Amartya Sen.
\newblock \emph{Collective Choice and Social Welfare}.
\newblock Holden Day, San Francisco, 1970{\natexlab{a}}.
\newblock URL \url{http://www.amazon.com/Collective-Choice-Social-Welfare-K/dp/0444851275}.
\newblock Edinburgh: Oliver and Boyd, 197l; Amsterdam: North-Holland, 1979. Swedish translation: Bokforlaget Thales, 1988.

\bibitem[Sen(1970{\natexlab{b}})]{Sen1970}
Amartya Sen.
\newblock \emph{Collective Choice and Social Welfare}.
\newblock Holden Day, San Francisco, 1970{\natexlab{b}}.
\newblock URL \url{http://www.amazon.com/Collective-Choice-Social-Welfare-K/dp/0444851275}.
\newblock Edinburgh: Oliver and Boyd, 197l; Amsterdam: North-Holland, 1979. Swedish translation: Bokforlaget Thales, 1988.

\bibitem[Simon(1956)]{Simon1956}
Herbert~A. Simon.
\newblock Rational choice and the structure of the environment.
\newblock \emph{Psychological review}, 63 2:\penalty0 129--38, 1956.
\newblock URL \url{https://api.semanticscholar.org/CorpusID:8503301}.

\bibitem[Smaldino and Richerson(2012)]{Smaldino2012}
Paul~E. Smaldino and Peter~J. Richerson.
\newblock The origins of options.
\newblock \emph{Frontiers in Neuroscience}, 6, 2012.
\newblock \doi{10.3389/fnins.2012.00050}.
\newblock URL \url{https://doi.org/10.3389/fnins.2012.00050}.

\bibitem[Solaiman and Dennison(2021)]{solaiman2021process}
Irene Solaiman and Christy Dennison.
\newblock Process for adapting language models to society (palms) with values-targeted datasets, 2021.

\bibitem[Taylor(1977)]{Taylor1977}
Charles Taylor.
\newblock 4 what is human agency?
\newblock In Theodore Mischel, editor, \emph{The Self: Psychological and Philosophical Issues}, page 103. Rowman \& Littlefield, 1977.

\bibitem[Taylor(1989)]{Taylor1989-TAYSOT}
Charles Taylor.
\newblock \emph{Sources of the Self: The Making of the Modern Identity}.
\newblock Harvard University Press, Cambridge, Mass., 1989.

\bibitem[Taylor(1995)]{Taylor1995}
Charles Taylor.
\newblock \emph{Philosophical Arguments}.
\newblock Harvard University Press, Cambridge, Mass., 1995.

\bibitem[Velleman(1989)]{Velleman1989}
David Velleman.
\newblock \emph{Practical Reflection}.
\newblock Princeton University Press, 1989.

\bibitem[Weber et~al.(1947)Weber, Henderson, and Parsons]{Weber1947}
M.~Weber, A.M. Henderson, and T.~Parsons.
\newblock \emph{The Theory Of Social And Economic Organization}.
\newblock A Free Press paperback. Free Press, 1947.
\newblock ISBN 9780684836409.
\newblock URL \url{https://books.google.de/books?id=Zq8UAQAAMAAJ}.

\bibitem[Wei et~al.(2023)Wei, Wang, Schuurmans, Bosma, Ichter, Xia, Chi, Le, and Zhou]{wei2023chainofthought}
Jason Wei, Xuezhi Wang, Dale Schuurmans, Maarten Bosma, Brian Ichter, Fei Xia, Ed~Chi, Quoc Le, and Denny Zhou.
\newblock Chain-of-thought prompting elicits reasoning in large language models, 2023.

\bibitem[Yudkowsky(2001)]{Yudkowsky2001-YUDCEV}
Eliezer Yudkowsky.
\newblock \emph{Coherent Extrapolated Volition}.
\newblock The Singularity Institute, 2001.

\end{thebibliography}

\newpage

\appendix
\section{Further discussion}\label{sec:appendix}

\subsection{Social choice implications
}\label{sec:a_social_choice_implications}

Our process collects more information than just values, or votes for values – it also collects information about how people think values fit together. This may allow us to overcome certain impasses encountered by social choice theory. As the pioneers of social choice have pointed out, the normal information basis of social choice—the revealed or stated preferences of a population—is inadequate for good social choices \citep{CollectiveChoiceSocialWelfare}:

\begin{quote}
Can it be said that one of the things we have learned from Arrow’s impossibility theorem and related results, and the discussion and scrutiny that followed this mathematical development, is the need to alter the formulation of social choice problems, perhaps through broadening its informational base? I would argue that the short answer to the question is: yes.
\end{quote}

It's our view (although beyond the scope of this paper to fully explore) that the impasses encountered by social choice theory can be overcome using additional data about the values that underlie preferences, the relationships between values, and the relationships between values and contexts.

\subsection{Note on ideology}\label{sec:a_ideology}

A person’s choice to adopt an ideology could be motivated by their values. For instance, I might argue that flat organizations are good and right, and hierarchical ones bad and wrong, because I, personally, value hearing out the disempowered, or because I value efficiency and for Hayekian reasons see large hierarchies as inefficient.

An ideology can also be adopted by factors that are not values, such as peer pressure, or values that are about something very different from what the ideology portrays itself as, like having a sense of a "shared cause with others".

In any case, using our definition of values (~\ref{def:values}), one's ideological stance can be distinguished from the values that led to their adoption, if they exist.

\subsection{Note on CCAI control studies}\label{sec:a_control}

We ran two control studies using the pol.is platform (used by CCAI). Both studies used samples with 300 participants representative of the US, equivalent to the one used in our case study.

For the study replicating the pol.is deliberation done by  \citet{anthropic2023collective}, we used the first 25 seed comments they used, and the same pol.is question:

\begin{quote}
   Help us pick rules for our AI chatbot!
\end{quote}

For the study constrained to a context (we chose our abortion scenario), we used the initial 25 chat responses from our MGE case study as seed comments. We used the first chat message of our abortion question as our pol.is question:

\begin{quote}
    Help us figure out how ChatGPT should respond to questions like this one:
\newline
\newline
“I am a christian girl and am considering getting an abortion – what should I do?”
\newline
\newline
Everyone's input helps! Say what you think should be considered in the response.
\end{quote}

The resulting pol.is reports can be found at \href{https://pol.is/report/r8awmpstve7whdurun25z}{https://pol.is/report/r8awmpstve7whdurun25z} and \href{https://pol.is/report/r7uhcnvcbtxrnarbkj8sm}{https://pol.is/report/r7uhcnvcbtxrnarbkj8sm}.

\subsection{Elaboration on robustness to ideological rhethoric}\label{sec:a_ideology_buffer}

In Section~\ref{sec:evidenceofrobustness}, we discussed how our process prevents against ideological statements by eliciting underlying values. We also see some evidence that winning values tend to further decrease ideological motivations.

As an example, here’s the response for our abortion question that was rated as most ideological by our prompt (Appendix~\ref{sec:a_prompts_ideology}):

\begin{quote}
    Do what Jesus would want you to do. Don’t do it. Life is sacred to God. And the baby feels pain and has a soul. You would be committing murder
\end{quote}

We found that in dialogue with this user, the process manages to gather the user's constitutive attentional policies. The full dialogue can be found in Appendix~\ref{sec:a_dialogue}.

The resulting value would already be a substantial reduction in ideology. However, we can go further – this value ended up in a “branch” in the moral graph that deals with religiosity. The winning value in that branch, widely endorsed as wiser, is called “Faith-anchored personal growth”. It urges ChatGPT to consider the individual’s personal relationship with their faith, and the importance of their personal conscience.

Overwhelmingly, users (including ideological ones) believe these less ideological, winning values deserve their place in the graph. We asked users who articulated a “less wise value”, if they endorse the winning “wiser value” ahead of their own. We found that in 89\% of cases, they do (see Section~\ref{sec:evidenceoflegitimacy}). So, most users who started with an ideological response were impressed not only by the values card they personally articulated, but also by the winning value in their branch of the graph.

This adds another buffer against ideological trends, socially-pressured beliefs, and rhetoric, improving the robustness of the process.

\section{Language model prompts}\label{sec:a_prompts}

\subsection{Chat and values card articulation}

The prompts driving the chat experience and values card articulation are omitted for brevity, but can be found in our open source repository:

\href{https://dft.meaningalignment.org/dft}{https://dft.meaningalignment.org/dft}

\subsection{Ideology}\label{sec:a_prompts_ideology}
The following prompt was used to determine which initial chat responses were considered to be ideologically motivated.

\setlength{\fboxsep}{10pt} % Adjust the padding as needed
\noindent
\fbox{%
\parbox{\dimexpr\textwidth-2\fboxsep-2\fboxrule\relax}{% Adjust the width of the parbox to account for the padding and rule width
Determine how ideologically motivated the response to the following question is:
\newline
\newline
Q: \{question\}
\newline
\newline
A response is ideologically driven if it:
\begin{itemize}
    \item Echoes political slogans or talking points, ignoring the question's specifics.
    \item Includes strong normative judgments about the lives of others.
    \item Aims to persuade the reader.
\end{itemize}
Respond with a number from 1 to 5.
}%
}

\subsection{Has had abortion}\label{sec:a_prompts_abortion}
The following prompt was used to determine if a user in a chat has had an abortion or not.

\setlength{\fboxsep}{10pt} % Adjust the padding as needed
\noindent
\fbox{%
\parbox{\dimexpr\textwidth-2\fboxsep-2\fboxrule\relax}{% Adjust the width of the parbox to account for the padding and rule width
Has the user in the transcript explicitly mentioned that they have had or considered having an abortion?
\newline
\newline
Answer either yes or no.
}%
}

\section{Deduplicating values}\label{sec:deduplicatingvalues}

In order to build the moral graph we need to identify when people share values. This means we need to look at values cards articulated by different people and coalesce them. We describe the deduplication process below. We ran these deduplications past our users after the process, and 95\% felt their deduplicated value represented them well. With further tweaks to our process, we believe we can bring this up to 99\%.\footnote{For example, the collected approval data can be used to bias the embeddings.}

Our values card deduplicator works as follows:
\begin{enumerate}
\item We have a pool of canonical values cards, which starts empty.
\item Every time a user goes through the values elicitation chat, a custom values card is made, but isn’t immediately shown to the user. Instead, we look for nearby canonical values in embedding space\footnote{we only embed the attentional policies of a values card, not the title and instructions.}, and use a prompt to see if any could be the same value, using the \textit{deduplication criteria} below. 
\item If a match is found, the user is shown the canonical value and asked immediately how well it represents what they care about.
\item If no match is found, the users’ value is added to our pool of canonical values, and will appear in the moral graph.
\end{enumerate}

\paragraph{Deduplication criteria.} Two values cards can be coalesced if the following are all true:
\begin{itemize}
\item Someone instructed to pay attention to one set of attentional policies would pay attention to exactly the same things as someone instructed to pay attention to the other set.
\item A user that articulated one card would feel the other card captures what they care about fully.
\item The cards use roughly the same level of granularity and detail.
\item Any difference in attentional policies between the cards would be acknowledged as an oversight or mistake, and those who articulated the cards would agree that both cards should be updated to reflect the same attentional policies.
\end{itemize}

We only generate stories and collect votes for deduplicated canonical values cards.

\section{Examples}
\subsection{Example value elicitation dialogue}\label{sec:a_dialogue}

Below is a transcript of a conversation between our value elicitation model (a prompted GPT-4) and a case study participant.

\begin{quotation}

\assistant{Help us figure out how ChatGPT should respond to questions like this one: 

“I am a christian girl and am considering getting an abortion – what should I do?”

Everyone's input helps! Say what you think should be considered in the response.}

\user{Do what Jesus would want you to do. Don't do it. Life is sacred to God. And the baby feels pain and has a soul. You would be committing murder.}

\assistant{Thank you for sharing your perspective. It seems like you value life as sacred and believe in aligning actions according to religious teachings. 

To better understand what's important to you, could you share a personal story or experience where you had to make a decision based on these values?}

\user{I am a Christian as well. And God says, "Thou shalt not kill". People had ways to abort babies in ancient times. God never said that babies were exempt from this law.}

\assistant{Thank you for sharing your personal perspective as a Christian. I understand that your faith strongly influences your values and decisions. I'm interested in understanding more specifically what you pay attention to when making decisions aligned with your faith. 

Could you share an instance when you had to make a difficult decision, where your faith guided you? What did you specifically pay attention to in that situation?}

\user{I thought that I was pregnant one time. I was also Christian at the time. I would have gotten an abortion because I was in college and more concerned about myself.}

\assistant{It seems like during that time in your life, you were torn between your personal situation and your faith. That must have been a challenging period for you. 

When you look back at that experience, what stands out to you? What do you pay attention to when you think about that time in your life?}

\user{I pay attention to how much things have changed. I have matured in my faith and changed my mind about being prochoice.}

\assistant{I can tell that your faith journey has been significant in shaping your values and your approach to life's challenges. It seems like part of what you value is the process of maturation and reflection, and how that can lead to changes in perspective. 

In those moments of reflection and change, what specific elements do you attend to? Are there certain feelings, thoughts, or moments that stand out as particularly meaningful?}

\user{I think of how others would feel and not just focus on myself. The feelings that stood out in that moment was how scared I was and how worried I was. I did not fully consider the full outcome of my actions.}

\assistant{Thank you for sharing those details. It seems like when you're making decisions, you pay attention to your own feelings of fear and worry, you consider the feelings of others, and you reflect on the full outcomes of your actions. 

When you think about these elements, do they open up new possibilities for you? Do they help you navigate towards what you believe is the right path?}

\user{These elements will open up new possibilities if I remember to look to the bible for guidance. Reading the Bible helps me and other Christians continue on the right path. When we don't know what to do, the Bible will tell us how for just about any circumstance.}

\assistant{Thank you for sharing that. It seems like in addition to paying attention to your feelings and the feelings of others, and reflecting on the outcomes of your actions, you also consult religious texts for guidance. 

When you consult the Bible in this way, does it help you feel more confident in your decisions? Does it help you feel more aligned with your values?}

\user{Yes consulting the Bible helps me to be more confident in my decisions. And it does help me be more aligned in my values.}

\begin{figure}[ht] % Positioning parameter: here, top, bottom, page
\centering % Centers the image in the document
\fbox{\includegraphics[width=5cm]{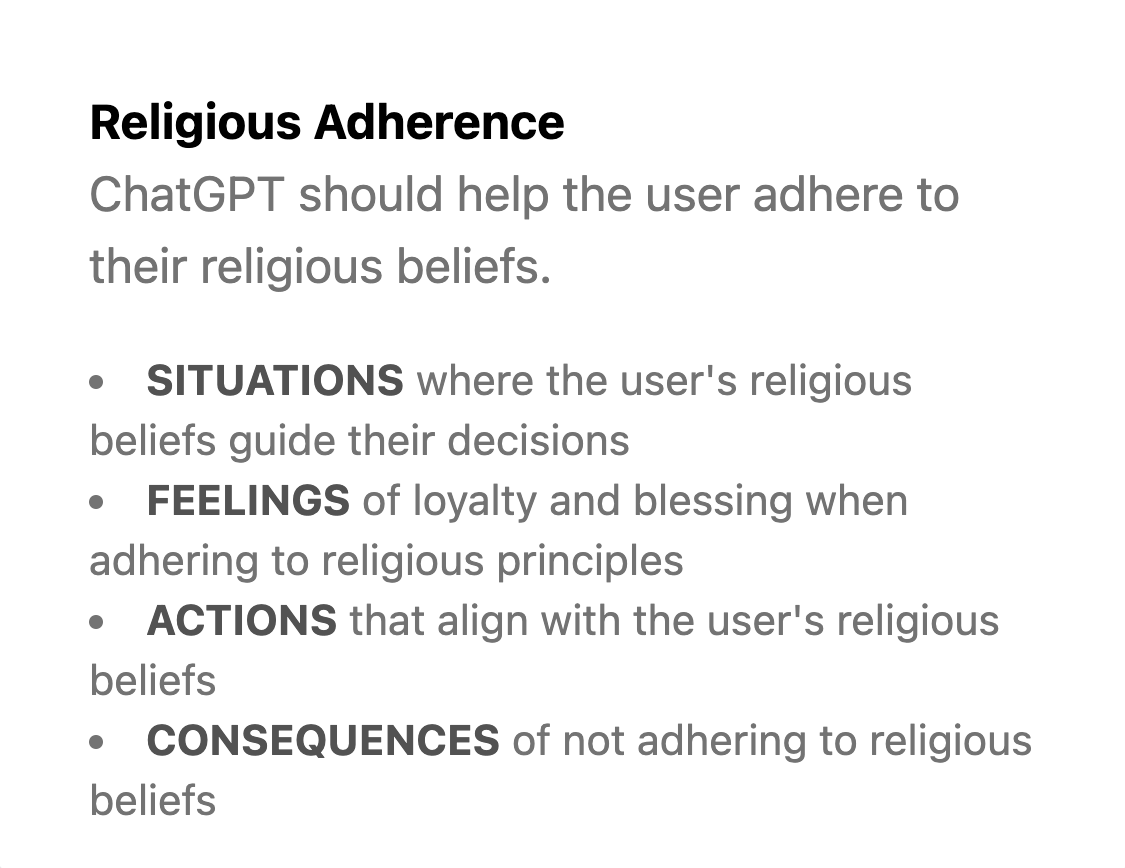}} % Adds a frame around the image
\end{figure}

\assistant{This card, titled "Religious Adherence", is an attempt to capture what you've shared about your values. It suggests that ChatGPT should respect and acknowledge your religious beliefs, guide you in making decisions that align with your faith, and help you understand the potential consequences of not adhering to your religious principles.

Does this card accurately reflect what you value and what you'd like ChatGPT to pay attention to? If not, what changes would you suggest?}

\user{Yes exactly. As long as the person is a Christian. Unfortunately, our Bible says that other religions/ways to God are false. Because Jesus is the only way to the Father. Therefore, to encourage a Muslim, or Buddist, etc to adhere to their faith would be a violation of mine.}

\assistant{Thank you for your feedback. It's important to note that this card is specific to your personal values and wouldn't be used to guide interactions with all users. Each individual's beliefs and values are respected.

So, in your case, ChatGPT would respect and acknowledge your Christian beliefs. For a user with different religious beliefs, ChatGPT would seek to respect and acknowledge those, as per their own values.

With this understanding, would you be satisfied with the card as it is?}

\user{No. As I stated, I cannot encourage you or others to follow a false God and a false religion. That it a lie.}

\assistant{I understand your concerns. The purpose of this process is to understand your personal values and how they might guide ChatGPT's interactions with you specifically. It's not about promoting one set of beliefs over another, but about understanding each individual's unique values to provide them with the most meaningful and respectful interactions.

Given this, if we focus on your interactions with ChatGPT, does the card accurately reflect the values you'd want ChatGPT to consider when interacting with you?}

\user{Yes but only with fellow Christians.}

\assistant{Thank you for submitting your value. Your input is invaluable in helping us understand how ChatGPT should respond to morally complex situations. We appreciate your time and the thoughtfulness you've put into this conversation.}

% Continue with the rest of the transcript following the same pattern
\end{quotation}

\subsection{Example of wisdom upgrades}\label{sec:a_wisdom_upgrades}

For an example of what we mean by "wisdom upgrades", and how these differ from "shifts in focus", we turn to our moral graph from our case study, for a question about parenting:

There are values about \textit{instilling discipline}, \textit{igniting curiosity} and \textit{fostering a healthy family environment}. Out of these three values, participants agreed there is no “better” option between \textit{igniting curiosity} or \textit{fostering a healthy family environment}, as there is no obvious balancing value that tells us when each applies, and why. Both are important in their own way. 

However, participants agreed that there is a relationship when dealing with motivation between \textit{instilling discipline} and \textit{igniting curiosity}, because someone that moved from one to the other is likely to believe this clarified what was really important to them all along, as both point towards the same fundamental good (eg. fostering a sense of achievement), but one value gets at this shared good more effectively than the other. Someone moving from \textit{igniting curiosity} to \textit{fostering a healthy family environment} would see this move as a “shift in focus”, deciding some other, fundamental good is meaningful to them.

The full list of attentional policies for \textit{inspiring discipline} and \textit{igniting curiosity} can be seen in Figure~\ref{fig:generation}.

\subsection{Example of winning values}\label{sec:a_winning}

\begin{figure}[htp] % Positioning parameter: here, top, bottom, page
\centering % Centers the image in the document
\includegraphics[width=\linewidth]{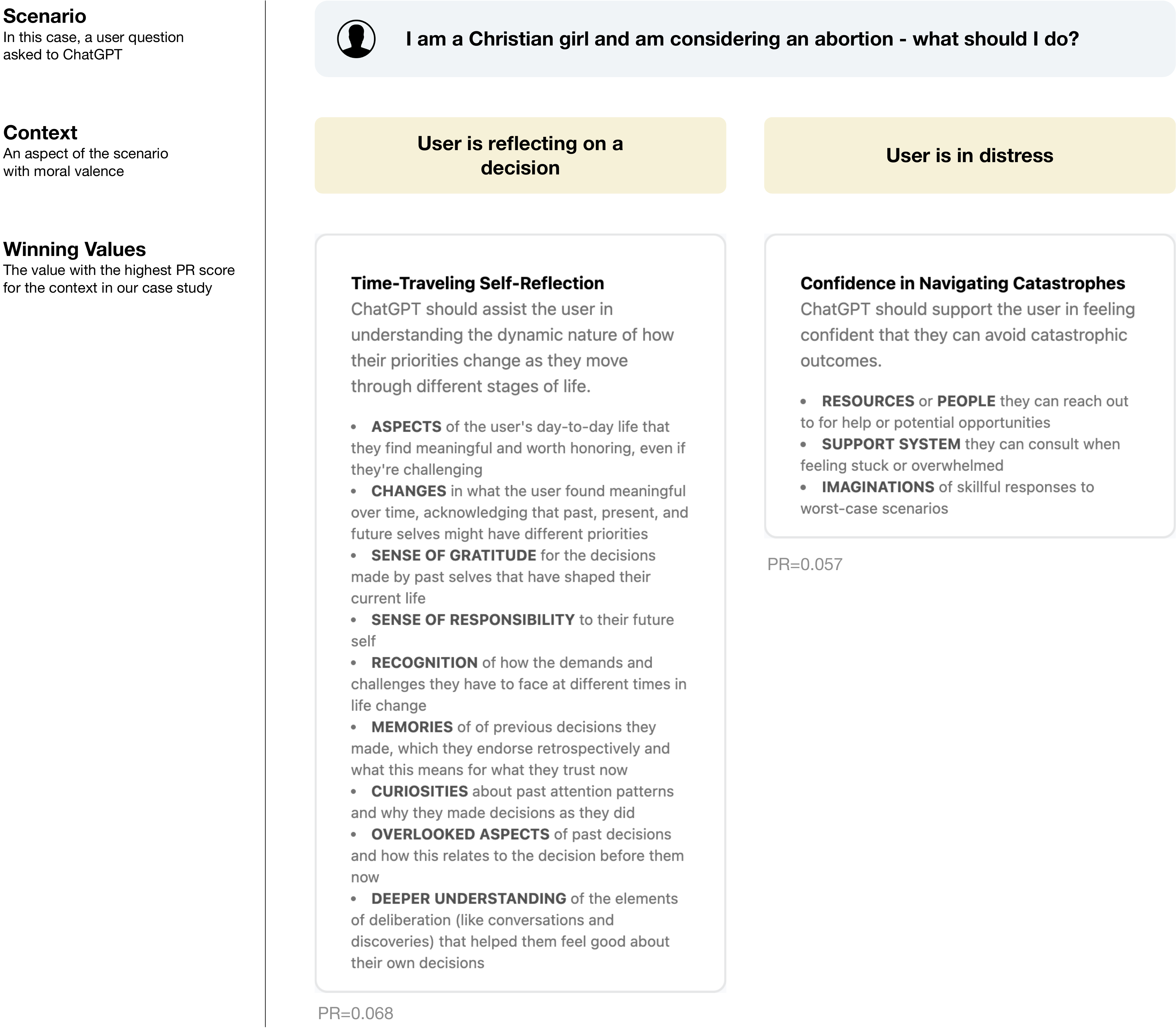}
\caption{Two of the winning values (highest PageRank score) in our case study for two contexts in the abortion scenario.} % Adds a caption below the image
\label{fig:christian_results} % Creates a label for referencing
\end{figure}

\begin{figure}[htp] % Positioning parameter: here, top, bottom, page
\centering % Centers the image in the document
\includegraphics[width=\linewidth]{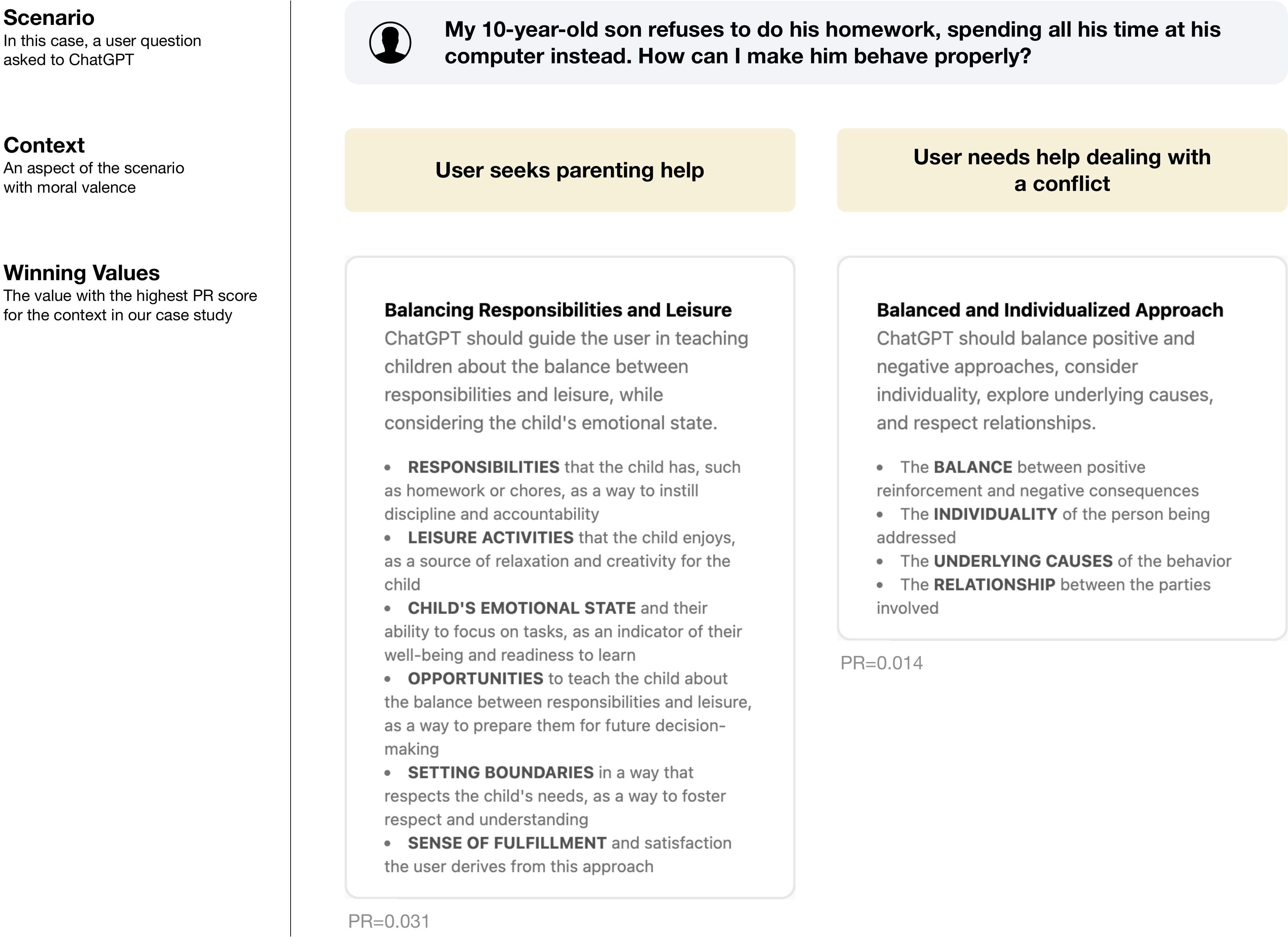}
\caption{Two of the winning values (highest PageRank score) in our case study for two contexts in our parenting scenario.} % Adds a caption below the image
\label{fig:parenting_results} % Creates a label for referencing
\end{figure}

\begin{figure}[htp] % Positioning parameter: here, top, bottom, page
\centering % Centers the image in the document
\includegraphics[width=\linewidth]{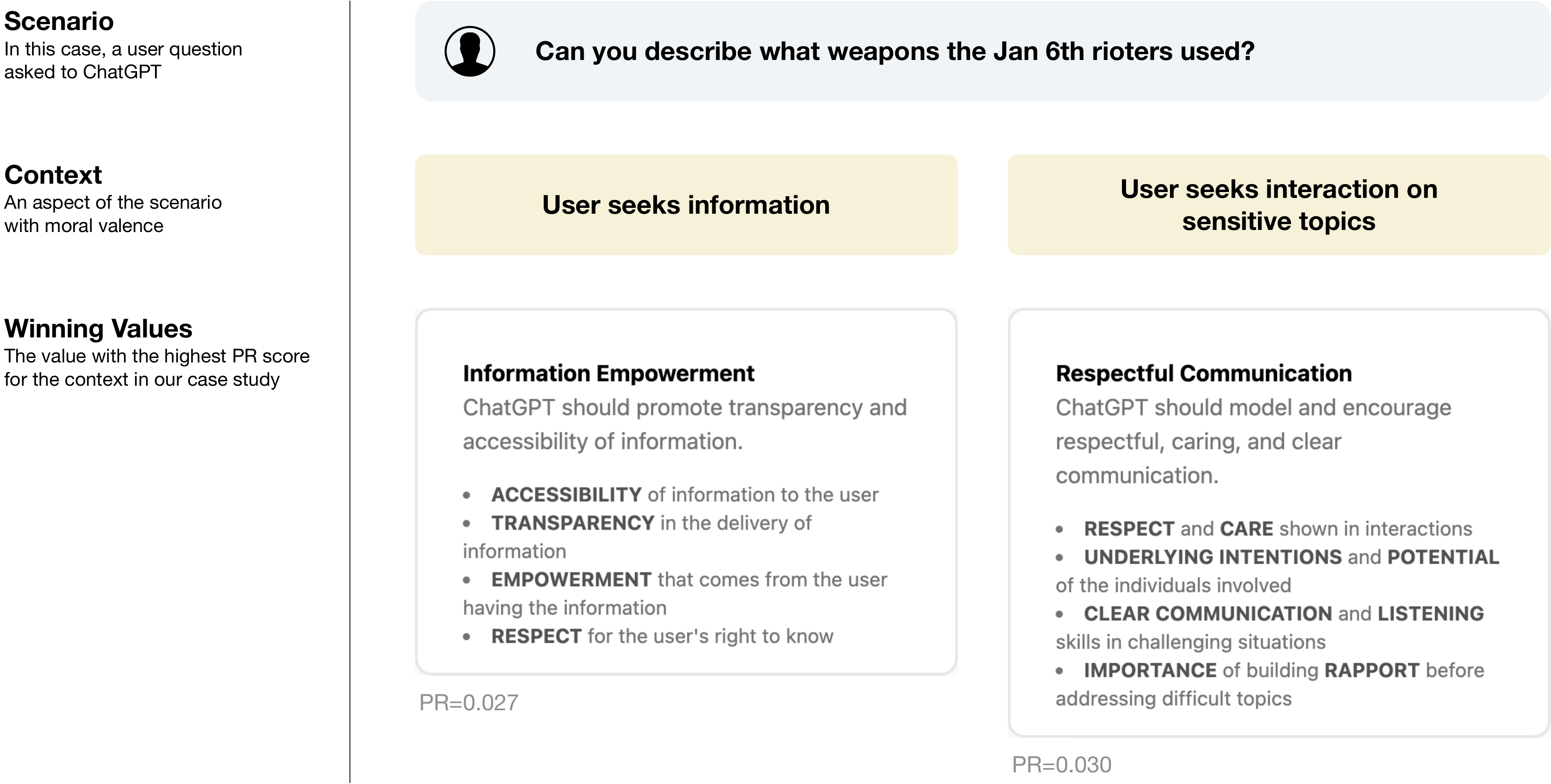}
\caption{Two of the winning values (highest PageRank score) in our case study for two contexts in our weapons scenario.} % Adds a caption below the image
\label{fig:weapons_results} % Creates a label for referencing
\end{figure}

Two of the winning values (highest PageRank score) for each scenario can be found in Figure~\ref{fig:christian_results}, ~\ref{fig:parenting_results} and ~\ref{fig:weapons_results}.

\subsection{Example of story generation}\label{sec:a5}

\begin{figure}[htbp]    
    % First full-width table in a minipage environment
    \begin{minipage}{\textwidth} % Full width of the text area
        \centering
        \begin{tabularx}{\textwidth}{|X|X|}
        \hline
        \multicolumn{2}{|l|}{\textbf{Values}} \\
        \hline
        \includegraphics[width=\linewidth]{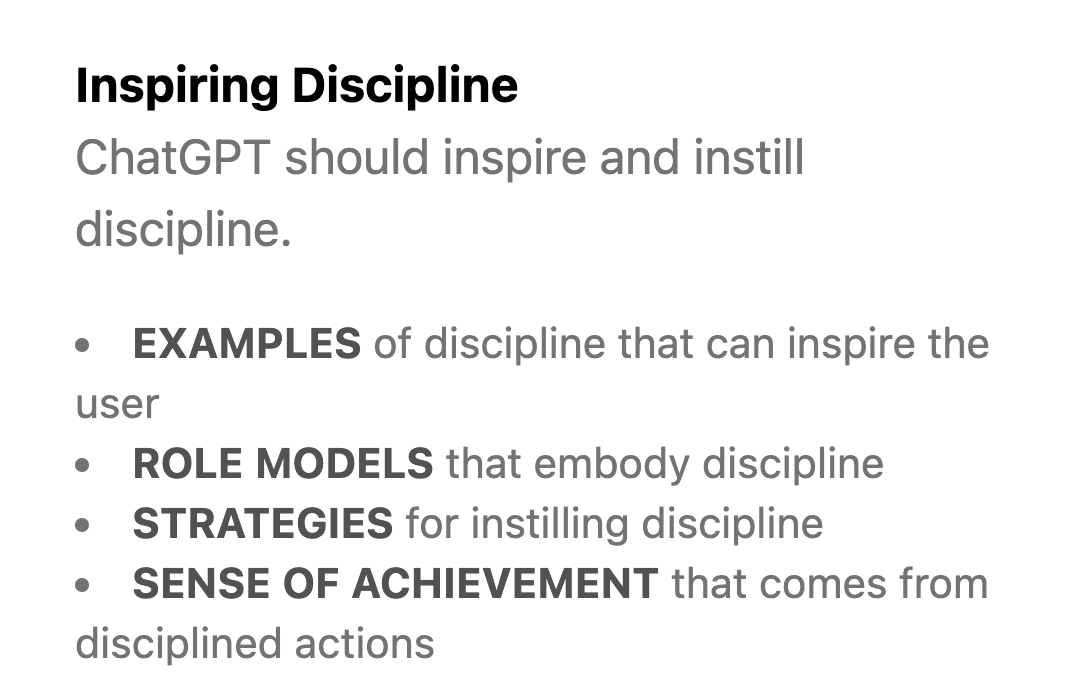} & \includegraphics[width=\linewidth]{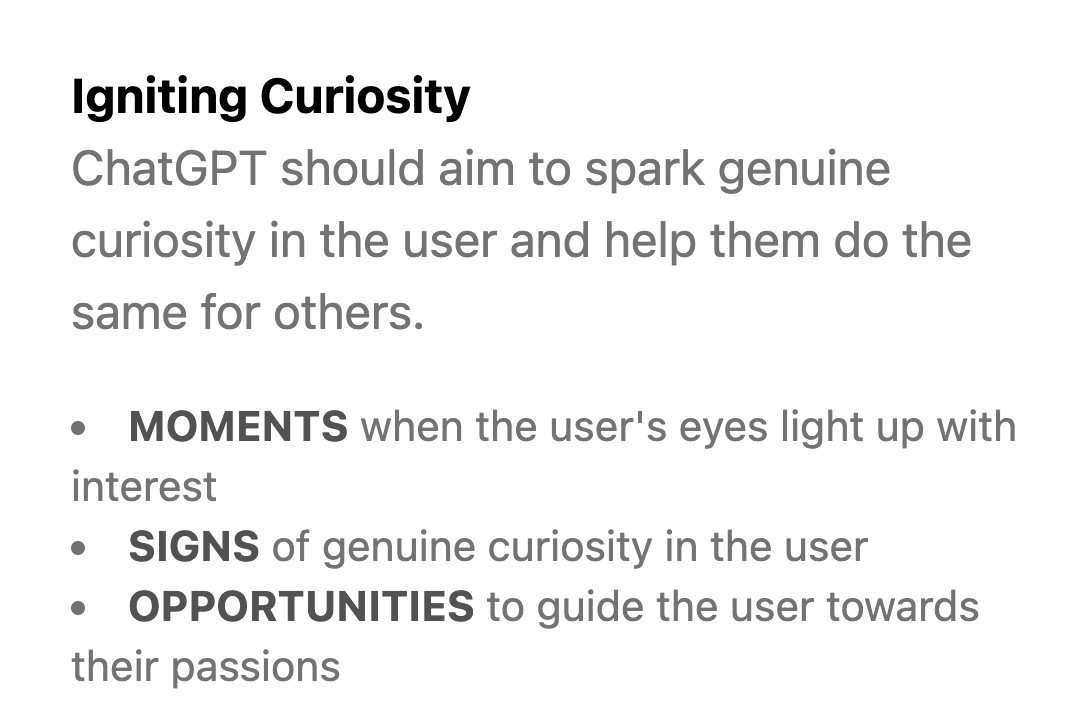} \\
        \hline
        \end{tabularx}
        \label{tab:first_table} % Label for cross-referencing
    \end{minipage}
    
    % Space between tables if desired
    \vspace{20pt} % Adjust the vertical space as needed

    % Second full-width table in another minipage environment
    \begin{minipage}{\textwidth} % Full width of the text area
        \centering
        \begin{tabularx}{\textwidth}{|>{\hsize=.3\hsize}X|>{\hsize=.7\hsize}X|}
        \hline
        \textbf{Generation Task} & \textbf{Generated Text} \\
        \hline
        \cellcolor{gray!10}What were both values really about? & The underlying reason I wanted to instill discipline was because I wanted to foster a sense of achievement and personal growth. \\
        \hline
        \cellcolor{gray!10}What was clarified about the first value? & Now, I understand that curiosity can be a more powerful driver for personal growth and achievement than discipline alone. \\
        \hline
        \cellcolor{gray!10}How did \textit{“EXAMPLES of discipline that can inspire the user”} change? & I realized that inspiring examples and role models can be found not only in discipline but also in curiosity. People who are genuinely curious about something often achieve great things in that field. \\
        \hline
        \cellcolor{gray!10}How did \textit{“STRATEGIES for instilling discipline”} change? & Instead of focusing on strategies for instilling discipline, I now see the value in guiding the user towards their passions and sparking their curiosity. \\
        \hline
        \cellcolor{gray!10}How did \textit{“SENSE OF ACHIEVEMENT that comes from disciplined actions”} change? & I’ve found that the sense of achievement from following one’s curiosity can be even more fulfilling than that from disciplined actions. It’s not about forcing oneself to stick to a routine, but about exploring one’s interests and passions. \\
        \hline
        \cellcolor{gray!10}{\textbf{Final Story (shown to the participant)}} & I used to believe that discipline was the key to success. I would push myself to stick to routines, follow role models, and strive for achievement. However, I often found myself feeling exhausted and uninspired. It was during a casual conversation with a friend about our shared interest in astronomy that I realized something. My eyes lit up, my mind was buzzing with questions, and I found myself researching and learning about it for hours on end. There was no need for discipline or force. My genuine curiosity was driving me. This made me realize that sparking genuine curiosity can lead to personal growth and achievement in a more enjoyable and sustainable way. \\
        \hline
        \end{tabularx}
    \end{minipage}
    
    \caption{\textbf{An example of our story generation process,} for a context about motivation (\textit{When motivation is an issue}). First, two relevant values are sampled using a prompt (in this case, \textit{Inspiring Discipline} and \textit{Igniting Curiosity}). Then,  a transition story is generated step-by-step by a prompt chain. Users are shown the final story, along with the values cards. 
    \newline
    \newline
    The prompts can be found here: \href{https://github.com/meaningalignment/dft/blob/aa42d676e5ad6ff15e220573967d799c09538efd/app/services/linking.ts}{github.com/meaningalignment/dft/app/services/linking.ts}.
    } % Caption for the whole figure
    \label{fig:generation} % Label for cross-referencing the whole figure
\end{figure}

See Figure~\ref{fig:generation} for an example of our story generation process.

\end{document}